\documentclass[twocolumn,showpacs,superscriptaddress, preprintnumbers , amsmath,amssymb,letterpaper]{revtex4-1}

\usepackage{dcolumn}
\usepackage{bm}
\usepackage{amsmath}
\usepackage{multirow}
\usepackage{graphicx}
\usepackage{float}
\usepackage[colorlinks=true]{hyperref}

\begin{document}


 \newcommand{\re}{\mathop{\mathrm{Re}}}
 \newcommand{\im}{\mathop{\mathrm{Im}}}
 \newcommand{\D}{\mathop{\mathrm{d}}}
 \newcommand{\I}{\mathop{\mathrm{i}}}
 \newcommand{\E}{\mathop{\mathrm{e}}}
 \newcommand{\unite}[2]{\mbox{$#1\,{\rm #2}$}}
 \newcommand{\myvec}[1]{\mbox{$\overrightarrow{#1}$}}
 \newcommand{\mynor}[1]{\mbox{$\widehat{#1}$}}
 \newcommand{\rmsemit}{\mbox{$\tilde{\varepsilon}$}}
 \newcommand{\mean}[1]{\mbox{$\langle{#1}\rangle$}}


\preprint{FERMILAB -PUB-2014-164-APC}
\date{\today}
\title{Ballistic Bunching of Photo-Injected Electron Bunches \\
with Dielectric-Lined Waveguides}

\author{F. Lemery} \affiliation{Northern Illinois Center for
Accelerator \& Detector Development and Department of Physics,
Northern Illinois University, DeKalb IL 60115,
USA} 
\author{P. Piot} \affiliation{Northern Illinois Center for
Accelerator \& Detector Development and Department of Physics,
Northern Illinois University, DeKalb IL 60115,
USA} \affiliation{Accelerator Physics Center, Fermi National
Accelerator Laboratory, Batavia, IL 60510, USA}

\begin{abstract}
We describe a simple technique to passively bunch non-ultrarelativistics ($\lesssim 10$~MeV) electron bunches produced in conventional photoinjectors. The scheme 
employs a dielectric-lined waveguide located downstream of the electron source to impress an energy modulation on a picosecond bunch. The energy modulation is 
then converted into a density modulation via ballistic bunching. The method is shown to support the generation of sub-picosecond bunch train with multi-kA peak currents.
The relatively simple technique is expected to find applications in compact, accelerator-based, light sources and advanced beam-driven accelerator methods. 
\end{abstract}

\maketitle

%
%
\section{Introduction}
Low-energy ($\lesssim 10$~MeV) electron beams are conventionally produced in photoemission electron sources based on radio frequency (RF) guns or ``photoinjectors". 
The final bunch length downstream of a photoinjector is dictated by the initial parameters including the photocathode-laser duration, transverse spot size and 
the electric-field amplitude in the  gun cavity and its phase relative to the laser. Typically, bunch lengths on the order of picoseconds are commonly produced in L- and 
S-bands RF guns. Shortening these bunches or producing trains of sub-ps microbunches is appealing to a variety of applications including 
ultra-fast electron diffraction~\cite{zewai,ued}, coherent accelerator-based, e.g., THz light sources~\cite{gover,mueller}, and injectors for short-wavelength 
advanced-accelerator concepts~\cite{aard, aard1}. \\

To date, bunch compression to produce kA peak current is often realized after acceleration to $\gtrsim 100$~MeV by employing dispersive sections arranged as, e.g.,  
magnetic chicanes~\cite{carlstenBC}. Alternative methods to shorten a relativistic bunch also include velocity bunching~\cite{velo1,velo2,velobunch,ferrario}, and ballistic bunching using 
an accelerating cavity operating at zero crossing. The latter method  demonstrated bunching at the sub-100-fs time scale~\cite{luiten} and could possibly 
produce shorter  temporal structures~\cite{klaus}. Similar methods have been extended to the mm-wave regime, e.g., by coupling laser-produced THz pulses 
to the beam using undulators~\cite{Thzbuncher}  or dielectric waveguides~\cite{franz}.

In addition, several techniques have demonstrated narrow-band THz radiation generation with photoinjector beams by coupling a density-modulated
bunch with electromagnetic-radiation mechanisms~\cite{muggli,yinebunch,piot,ychuang,boscolo}. Among these techniques, two of them are based on 
impressing a density modulation using a temporally-modulated photocathode-laser pulse~\cite{ychuang,boscolo,yuelinmod}. The use of such a
modulated laser was also experimentally shown to support the formation of  short-current spikes via wave breaking seeded by non-linear longitudinal 
space-charge effects~\cite{renkai}. 

Most recently, a technique to produce train of microbunches based on a dielectric-lined waveguide (DLW) was 
realized in a $\sim 70$-MeV accelerator~\cite{antipov4}. In the latter experiment a density modulation was produced using a small chicane to provide the  longitudinal 
dispersion necessary to convert the energy modulation imparted by the beam self-interaction with its short-range wakefield in the DLW structure.\\ 

In this paper, we propose a simple method extending the mechanism proposed in Ref.~\cite{antipov4} to low-energy beams. In our configuration a 
$\sim 5$-10~MeV ps-duration beam is energy-modulated as it passes through a DLW structure and ballistically bunched in a subsequent drift. 
Our approach is  similar to the bunching technique commonly used in klystrons \cite{buncher1,buncher2}. Owing to the low intrinsic energy spread typically 
achieved in photoinjectors, final beam currents in excess of kA's can be produced.

\section{Ballistic compression from wakefield-induced energy modulations\label{sec:scheme}}
A feature critical to the production of density modulated beams is the capability to produce the required large local longitudinal-phase-space (LPS) chirps via the self-wakefield in the considered DLW structure. We investigate this point with a cylindrically-symmetric DLW consisting of a hollow dielectric cylinder with inner and 
outer radii $a$ and $b$~\cite{rg}, and relative electric permittivity $\varepsilon_r$. The outer  surface of the dielectric is metalized.  We consider  the axial  longitudinal wakefunction 
modes supported by such a structure to be of the form~\cite{chao,stupakov}
\begin{eqnarray}~\label{eq:qe39rs}
w_{z,m} (\zeta )=\kappa_m \cos (k_m \zeta),
\end{eqnarray}
where $\zeta$ is the position of the observer charge referenced with respect to the source electron and $\kappa_m$ (with units of V/m/C) and $k_m$ are respectively the loss factor and 
wave vector associated to the $m$ mode supported by the DLW structure. The mode parameters $\kappa_m$ and $k_m$ are obtained following the  methodology  described in Ref.~\cite{rg} by 
numerically solving the dispersion equation. 

\begin{figure}[hhhhh!!!!!!!!!!!!]
\centering
\includegraphics[width=0.495\textwidth]{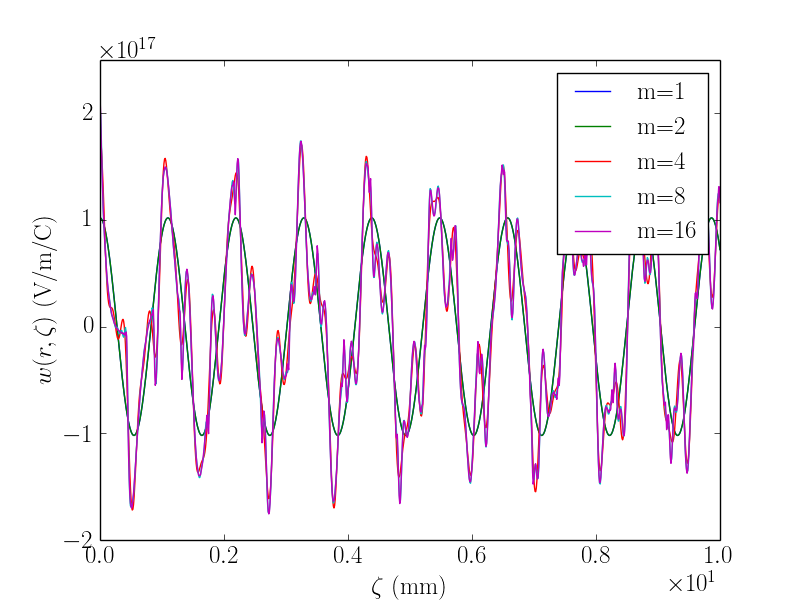} 
\caption{Wake function computed as $w_z(\zeta)=\sum_{l=1}^{m} w_{z,l}$ for $m=1$, 2, 4, 8, and 16 for a DLW structure with parameters $a=400$~$\mu$m,  $b=450$~$\mu$m,  and $\varepsilon_r=5.7$. 
The fundamental-mode (blue trace) wavelength is $\lambda_1\simeq 1.09$~mm. \label{fig:greens}}
\end{figure}

An example of computed Green's function for a structure with parameter $a=400$~$\mu$m,  $b=450$~$\mu$m,  and $\varepsilon_r=5.7$ (corresponding to 
diamond) appears in Fig.~\ref{fig:greens}. The Green's function converges after inclusion of 4 modes (the 50-$\mu$m thickness of the structure supports multiple modes with significant
axial fields).

Note that the field in Eq.~\ref{eq:qe39rs} and the wake function have no dependence on the transverse coordinates. The expected change in longitudinal momentum for a particle within and behind a bunch with  line-charge  distribution $\Lambda(z)$ is obtained from the convolution integral 
\begin{eqnarray}\label{eq:wkm0}
\Delta E (z) \simeq c \Delta p_z(z)= L_{dlw}\int_{-\infty} ^ {z} dz' \Lambda(z-z') w_z(z'),
\end{eqnarray}
where $L_{dlw}$ is the length of the DLW structure and $z$ the longitudinal coordinate within the bunch. 

\begin{figure}[t]
\centering
\includegraphics[width=0.485\textwidth]{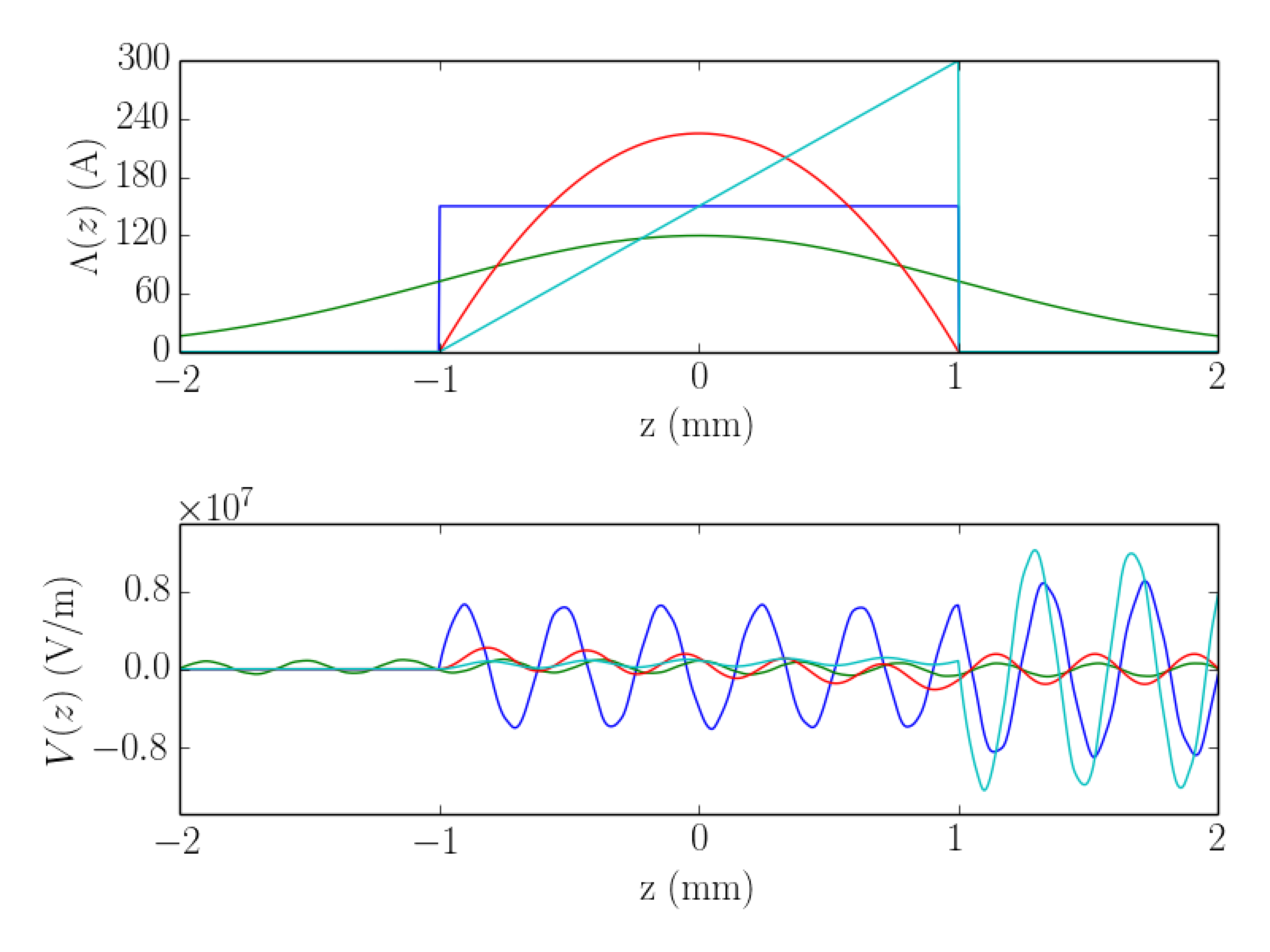} 
\caption{Charge distributions (top) and corresponding wake potential (bottom) for the same structure parameters as shown in Fig.~\ref{fig:greens} and for a 1-nC Gaussian bunch with variance (Gaussian distribution) or hard-edge half size (other distributions) of 1~mm.  The green, blue, red, and turquoise traces respectively correspond to the case of a Gaussian, parabolic, uniform, and linearly-ramped current distributions. The head of the bunch is at $z\le 0$. \label{fig:dist}}
\end{figure}

In contrast with an energy modulation imparted by external fields (e.g. from lasers or RF cavities), the modulation imparted via wakefield depends on the 
bunch shape. In particular, given the selected parameters for the DLW structure, one should ideally select an electron-bunch distribution with spectral contents 
capable of exciting the mode(s) supported by the structure; see Fig.~\ref{fig:dist}. 
Reference~\cite{antipov4} utilizes a relativistic bunch shape with linearly-ramped current profile as the associated energy modulation was  shown to exhibit a 
uniform-amplitude modulation throughout the bunch.

\begin{figure}[t]
\centering
\includegraphics[width=0.485\textwidth]{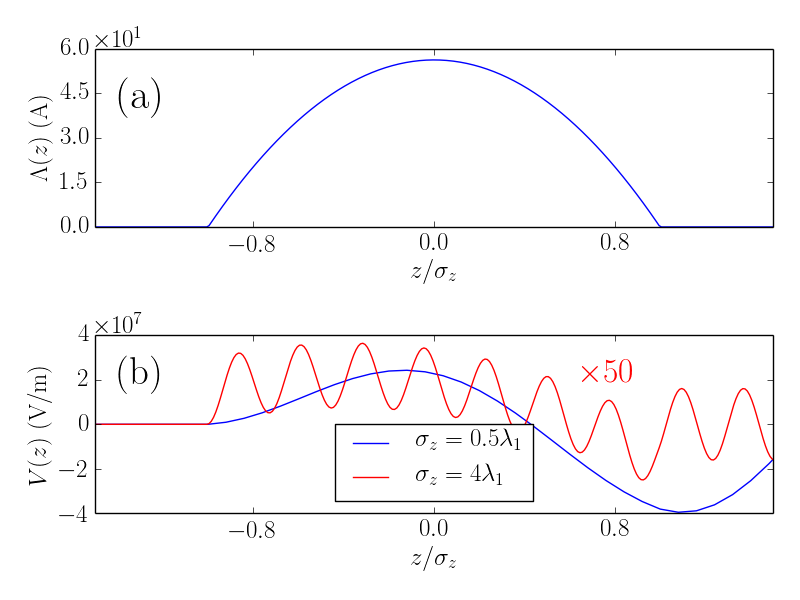} 
\caption{Charge distributions (a) and corresponding wake potential (b) for two cases of ratio between the rms bunch length $\sigma_z$ and fundamental-mode wavelength $\lambda_1$. The DLW structure parameters are identical to one used in Fig.~\ref{fig:greens}.  The head of the bunch corresponds to $z\le 0$.  The wake potential associated to the 
$\sigma_z=0.5\lambda_1$ case  is scaled by a factor 50 for clarity. \label{fig:wakeparabola}}
\end{figure}

In order to illustrate the proposed concept we elaborate a simple model based on  the ideal case of a line-charge electron bunch with a parabolic charge-density   
profile $\Lambda(z)=[3Q/(4a^3)](a^2-z^2)$
where $Q$ is the total bunch charge and $a$ the half width of the distribution; see Fig.~\ref{fig:wakeparabola}(a). The corresponding change in energy along the bunch is given by 
\begin{eqnarray}\label{eq:parawake}
\Delta E(z) &\simeq & \sum_{m=1}^{+\infty} {\cal E}  \left\{ \sin[k_m (z+a)] \right. \\ \nonumber
 && -  \left. k_m a \cos[k_m (z+a)] + k_m z \right\},
\end{eqnarray}
where ${\cal E} \equiv \frac{6\kappa_m L_{dlw}  Q}{4k_m^3 a^3}$. 
Considering only the fundamental mode ($m=1$) and assuming a ``cold" initial LPS with no correlation so 
that  $(z_i,\delta_i=0)$ (for all $i$), where $z_i$ and $\delta_i$ are respectively is the axial coordinate and fractional momentum spread associated to the $i^{th}$ electron.  
The final fractional momentum spread downstream of the DLW structure becomes  
\begin{eqnarray}~\label{eq:delta}
\delta_f (z_f) \simeq \frac{\Delta E(z_f)}{E_i+ Q \kappa_1 L_{dlw} a \mbox{sinc}(k_1 a)}, 
\end{eqnarray}
where $E_i$ is the bunch's initial mean energy, $z_f=z_i$, and $\mbox{sinc}(x)\equiv \sin(x)/x$ is the usual ``sampling function".

After a section with longitudinal dispersion $R_{56}$, the energy modulation induces a density modulation and the final longitudinal coordinate of an electron is mapped 
as $z_d=z_f+R_{56}\delta_f $ under a linear single-particle dynamics approximation.

We first consider the case when the root-mean-square (rms)  bunch length satisfies  $\sigma_{z,i}\equiv \mean{z_i^2}^{1/2}=a/\sqrt{5} \gtrsim \lambda_1 \equiv 2\pi/k_1$ so 
that an energy modulation along the bunch can be impressed; Fig.~\ref{fig:wakeparabola}(b, red trace). In such a case the second term in Eq.~\ref{eq:parawake} dominates the short-wavelength modulation 
structure and the final longitudinal coordinate is approximately given by 
\begin{eqnarray}
z_d \simeq  z_i - \frac{R_{56}  {\cal E} }{E_i+ Q \kappa_1 L_{dlw} a \mbox{sinc}(k_1 a) } \cos[k_1 (z_f+a)].
\end{eqnarray}

At the zero-crossing locations, i.e. the locations along the bunch $z_{f,n}$ such that $\delta_f(z_{f,n}) \propto \cos[k_1 (z_{f,n}+a)] =0$, the local LPS correlation is given by  

\begin{eqnarray}
{\cal C}\equiv \frac{d\delta_f}{dz_f}\bigg|_{z_{f,n}} \simeq \frac{k_1 {\cal E}}{E_i },
\end{eqnarray}
where we have further assumed that the $\mbox{sinc}(k_1 a)$ in the denominator is negligible. The maximum bunching occurs at these zero-crossing points when the following beamline provides a longitudinal dispersion $R_{56} = - \frac{1}{\cal C} $. 
The characteristic length of the microbunches formed is approximately given  by  $\sigma_z \simeq R_{56} \tilde{\sigma_{\delta}}$ where $\tilde{\sigma_{\delta}}$ is the 
uncorrelated (or slice) rms fractional momentum spread. The microbunches' separation is $\Delta z \equiv z_{f,n}-z_{f,n-1}\simeq \lambda_1$ for an incoming beam with vanishing  
correlated energy spread upstream of the DLW structure. \\

At relativistic energies, the  longitudinal dispersion $R_{56}$ necessary to form the microbunches is often provided by a dispersive section, e.g., a bunch-compressor chicane~\cite{carlstenBC} as accomplished in Ref.~\cite{antipov4}. Here we note that at energies below $\sim 10$~MeV (non-ultra-relativistic  regime), the large LPS slope resulting from the large axial fields supported in a DLW requires a relatively small $R_{56}$ that can be readily produced by a drift space. A drift with length $D$ has a longitudinal dispersion 
\begin{eqnarray}
R_{56} \simeq  -\frac{D}{\gamma^2}, 
\end{eqnarray}
where $\gamma$ is the bunch's Lorentz factor and we take $\beta \equiv  (1-1/\gamma^2)^{1/2} \simeq 1$ for simplicity. \\

Practically, for a $\sim 5$-MeV electron bunch passing through a 10-cm long DWL structure capable of supporting  $\sim 0.5$~MV/m peak field a ``local" chirp   ${\cal C} \simeq 10^{3}$~m$^{-1}$ can be obtained for a 0.5-mm modulation wavelength. The corresponding local density spike could form via  ballistic bunching after a drift of length below $D\leq 1$~m. The expected modulation amplitude $\sim 0.5$~MeV is much larger than the typical uncorrelated energy spread of a few keV routinely achieved in RF guns~\cite{huang,huening}. Additionally, the relatively low $R_{56}$ and small uncorrelated energy spread are also beneficial to the production of very short ($< 100$-fs) density spikes. 
This simple estimate motivates further investigation of the scheme using a bunch generated by a conventional photoemission electron gun. \\

In addition, furthering our point about the dependence of the energy modulation on bunch parameters we now examine the case when the rms bunch length fulfills $\sigma_{z,i} \simeq \lambda_1/2$; see Fig.~\ref{fig:wakeparabola}(b, blue trace). 
In this regime, the induced energy change along the bunch produced an energy depression between the head and tail of the bunch. The produced correlation between the depleted 
energy location and tail has the proper sign to be  compressed via ballistic  bunching.  Although the introduced chirp is nonlinear, it eventually can lead to the 
production of high-peak-current spikes. This approach however only bunches a fraction of the bunch and actually debunch the beam within the front of  the bunch. Despite this drawback, this scheme is 
appealing given its simplicity and absence of need for a precisely synchronized external field  as used in ballistic bunching using a buncher cavity~\cite{luiten}. This passive 
bunching method is therefore inherently self synchronized and in principle not subject to time jitter (the main source of jitter is associated to charge fluctuations that impact the imparted energy modulation and could consequently result in shot-to-shot fluctuation of  the peak-current value).

\begin{figure}[h!]
	\centering
	\includegraphics[width=0.475\textwidth]{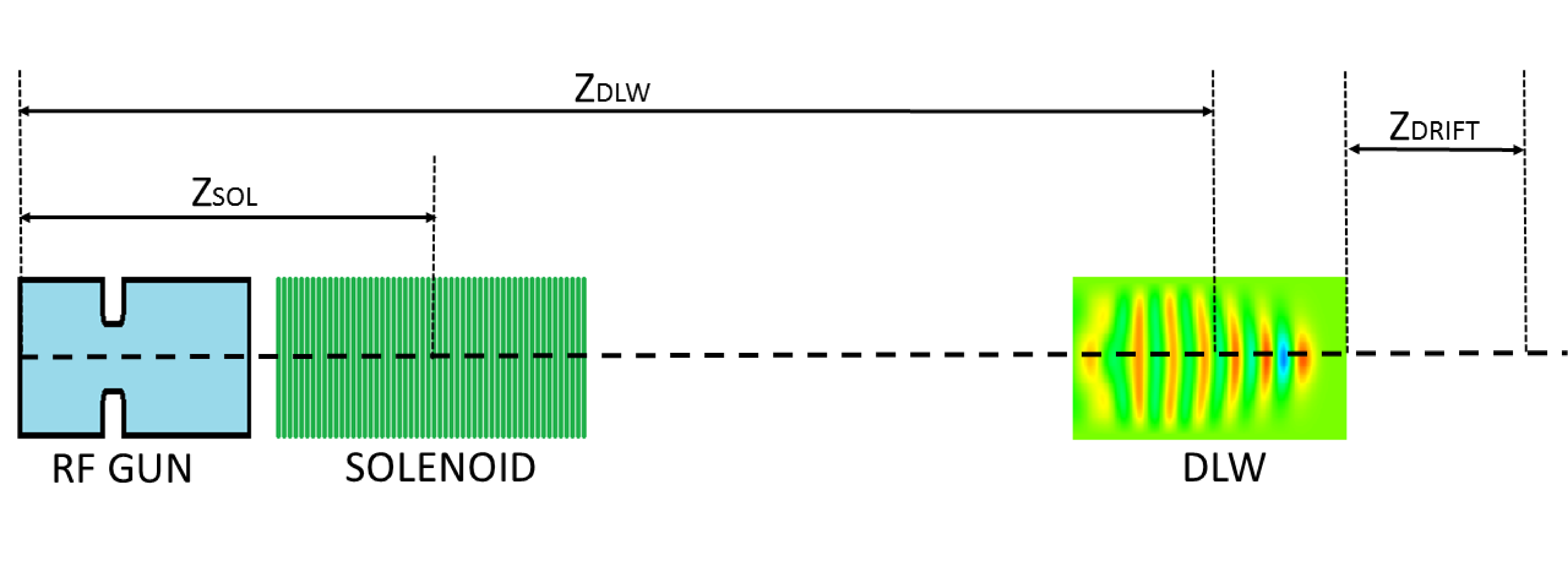} 
	\caption{Overview of the photoinjector setup used for the numerical simulations. The distances $Z_{SOL}$ and  $Z_{DLW}$ correspond respectively to the location of the center 
	of the solenoid and DLW structure referenced to the photocathode surface,  and $Z_{drift}$ represents the drift distance downstream of the DLW structure necessary for  ballistic   
	bunching. \label{fig:RFgun}}
\end{figure}

Finally, it should be pointed out that higher-order, e.g., dipole, modes can also affect the bunch transverse dynamics but are neglected in the present treatment as we assume the bunch is cylindrical-symmetric 
and is centered on the DLW axis.  Given the short length of the DLW considered in the remainder of the paper, possible detrimental effects on the transverse beam dynamics can 
be practically corrected, e.g., by mounting the DLW structure on translational stages. 

\section{Numerical modeling and analysis}

To explore the possibilities discussed in the previous section we perform beam-dynamics simulations. The numerical simulations are carried with 
the beam-dynamics program {\sc astra}~\cite{astra} which takes into account space-charge effects 
using a cylindrical-symmetric quasi-static space charge algorithm. The beam-DLW interaction is modeled via the Green's function approach briefly outlined above 
and detailed in Ref.~\cite{dohlus}. In the simulations the electron bunch is modeled as an ensemble of 100,000 macroparticles. 

To characterize the temporal structure of the bunch, we represent the macroparticles' temporal distribution as $\Lambda(z)=\frac{1}{N} \sum_{i=1}^N \delta(z-z_i)$ 
and compute the bunch  factor factor (BFF)  $\widetilde{F}(\omega) =|1/(2\pi) \int_{-\infty}^{+\infty} \Lambda(z/c) e^{-i\omega t}|^2$ as
\begin{eqnarray}
	\widetilde{F}(\omega) = \frac{1}{N^2} \left(\mathopen\bigg|\sum_i^N\cos \frac{\omega z_i}{c}\mathclose\bigg|^2 +  \mathopen\bigg|\sum_i^N\sin \frac{\omega z_i}{c}\mathclose\bigg|^2  \right), 
\end{eqnarray}
where $N$ is the number of macroparticles used in the simulation. The BFF is commonly used to characterize the performance of accelerator-based radiation 
source~\cite{saxon}. We note that in some cases, e.g. for the production of coherent radiation, transverse suppression effects might be prominent and should be properly 
accounted for by utilizing a three-dimensional expression for the BFF; see, e.g., Ref.~\cite{saldin}.  

\subsection{Sub-picosecond bunch train formation~\label{sec:train}}

We first investigate the practical realization of the scheme described in section~\ref{sec:scheme} to produce trains of sub-picosecond bunches and to demonstrate the 
versatility of the method, we consider two examples of implementation. The  generic setup consists of an RF-gun electron source followed by a DLW as diagrammed in 
Fig.~\ref{fig:RFgun}. Downstream of the DLW the beam is focussed with a second solenoid, e.g., to produce a waist at the location a transition-radiation target. 
The RF gun is taken to be an S-band (2.856~GHz) 1/2-cell cavity similar to the one currently in use at the linac coherent light source (LCLS)~\cite{slacgun}. Similar 
results are then confirmed using a 1/2-cell L-Band (1.3 GHz) gun similar to the one used at the FLASH facility in DESY~\cite{pitz}.  

\begin{table}[h]
	\caption{\label{tab:beamparam}
	Beamline settings and DLW-structure parameters used in the {\sc astra} simulations. The beamline configuration with some of the associated parameters is depicted 
	in Fig.~\ref{fig:RFgun}.}
	\begin{ruledtabular}
		\begin{tabular}{lcc r}
			  & S-Band & L-Band &  \\
			\hline
			  parameter &   &   & units \\
			\hline
			Laser pulse RMS duration  & 3 & 7 & ps\\
			Laser pulse rise time  & 100 & 100 & fs\\
			Laser RMS spot size & 0.72  & 1.1 & mm\\
			Initial charge 		& 1  & 1 & nC\\
			Peak field on cathode  & 120    & 34  & MV/m\\
			Solenoid 1 position	& 0.20      & 0.0  &m\\
			Solenoid 1 strength	& 0.26     & 0.17  & T\\ 
			Solenoid 2 position	& 1.35     & 1.0  & m\\
			Solenoid 2 strength	& 0.45    & 0.15&  T\\ 
			\hline
			DLW position		& 0.9     & 0.34&m\\
			DLW inner radius $(a)$ & 350    & 500  & $\mu$m\\
			DLW outer radius $(b)$  & 363   & 550  & $\mu$m\\
			DLW length  		& 11      & 4  & cm\\
			DLW fund. frequency $f_1$ & 1000    & 400  & GHz\\
			\hline
			Transmission through DLW		& 85 & 98  & \% \\
			Average kinetic energy		& 6.1   & 3.8 &  MeV\\
		\end{tabular}
	\end{ruledtabular}
\end{table}

The photocathode-laser distribution was chosen to follow a plateau temporal distribution and its transverse size along with the location of the DLW, and solenoid strength were 
optimized using a multi-objective optimizer~\cite{genOpt} to maximize beam transmission through the structure and minimize the transverse beam size at the DLW 
center. The list of optimized operating parameters are displayed Tab.~\ref{tab:beamparam} (``S-band" column). We note that the choice of the DLW parameters is a compromise 
between modulation wavelength $\lambda_1$, energy modulation amplitude -- which affects the bunching parameter -- and beam transmission.
For example, a shorter DLW structure relaxes the requirements on beam sizes and emittances at the structure, but necessitates a longer drift to bunch 
the beam (as the imparted energy amplitude modulation is smaller than for a longer structure). 
Additionally, the number of potential microbunches depends on the incoming bunch length and $\lambda_1$. For example, a Gaussian bunch with rms length $\sigma_z$ will eventually result 
in the generation of  $N_b \sim 4\sigma_z/\lambda_1$ microbunches; varying $\sigma_z$ for a given bunch charge and $\lambda_1$  affects the initial peak current and 
amplitude of the imparted energy modulation as inferred from Eq.~\ref{eq:delta}.

\begin{figure}[hhhh!!]
\centering
\includegraphics[width=0.46\textwidth]{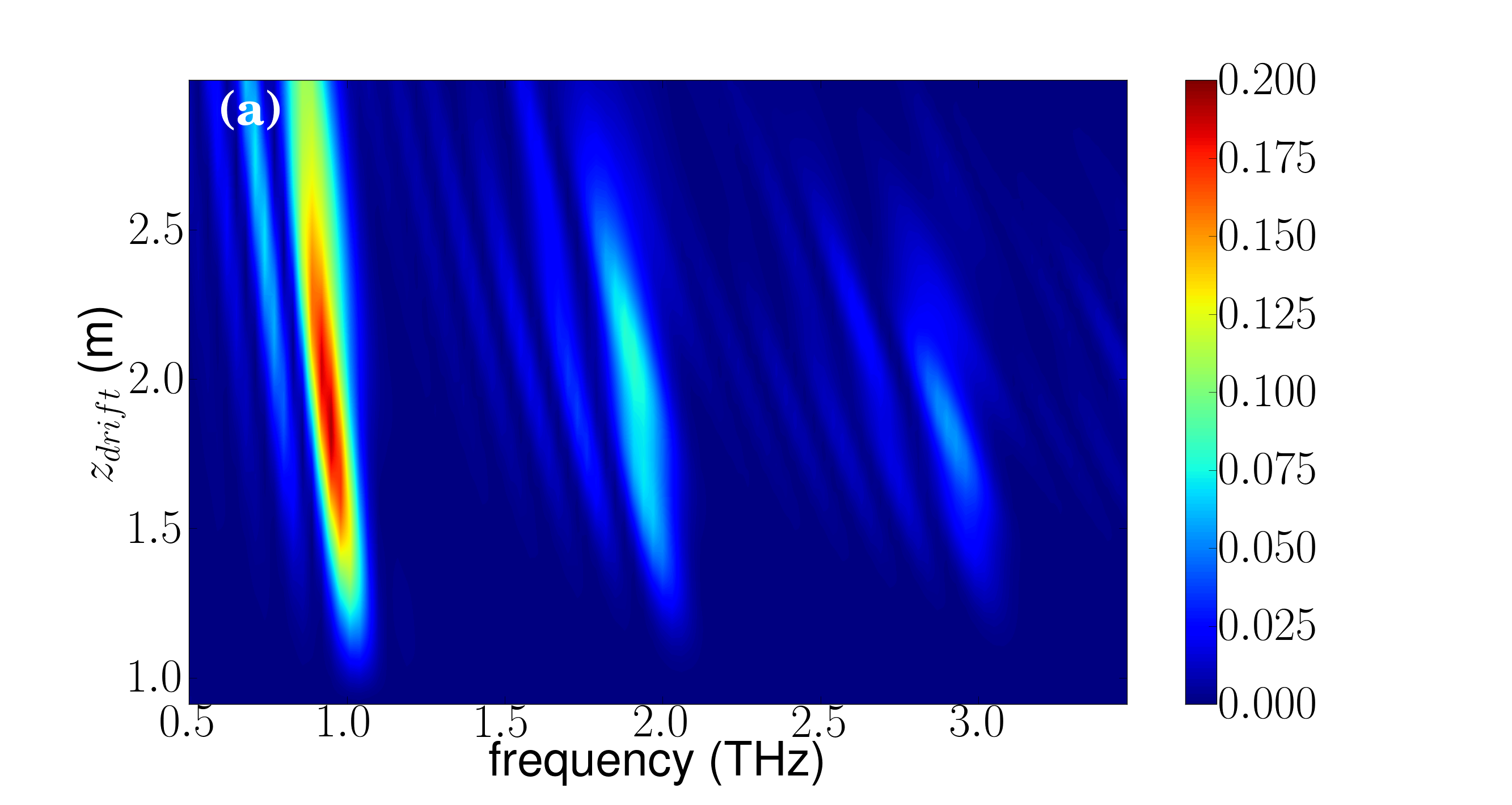} 
\includegraphics[width=0.46\textwidth]{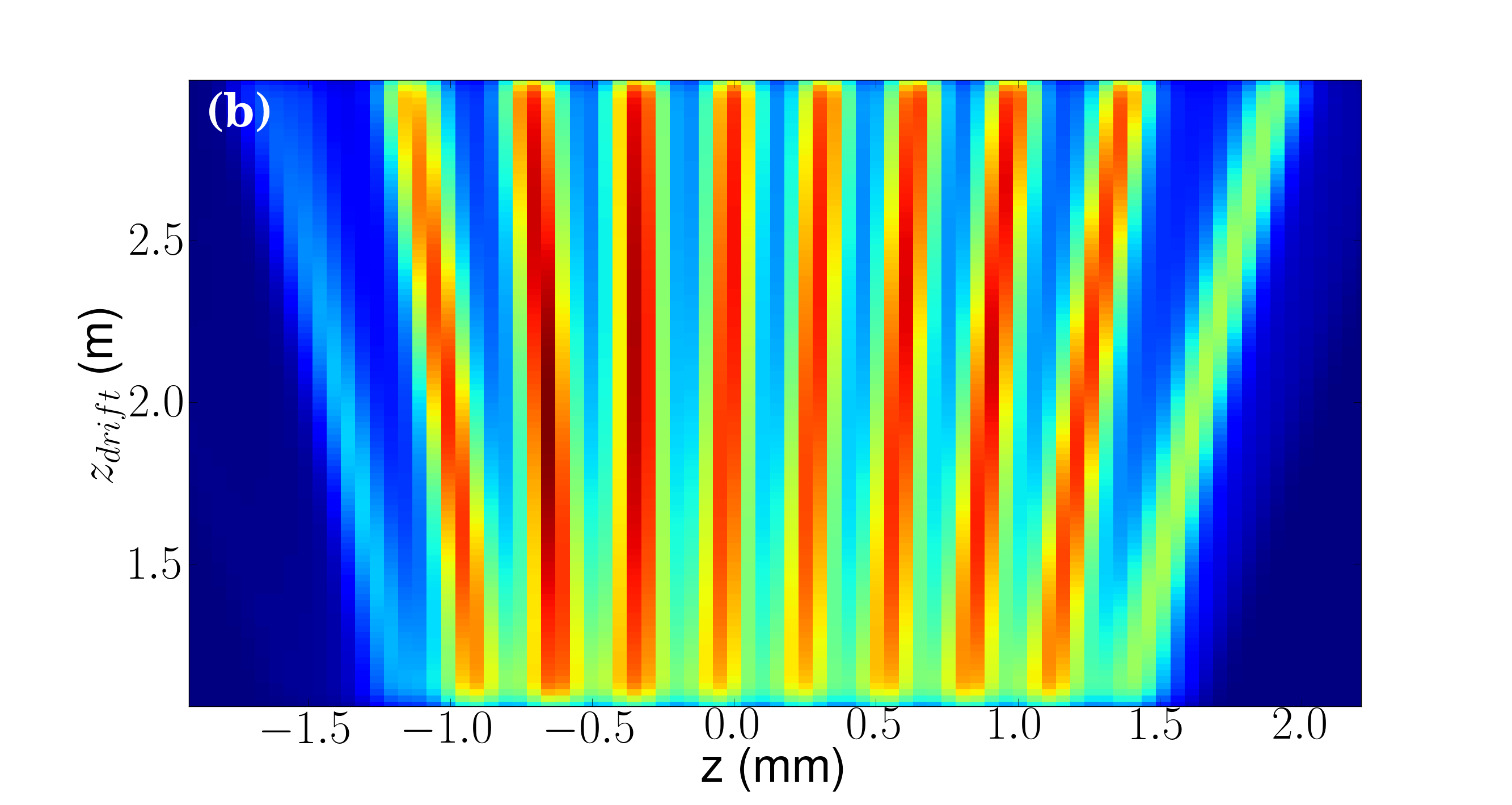}
\caption{Bunch form factor (BFF) (a) and bunch longitudinal density (b) evolution as a function of the drift length referenced with respect to the DLW exit.  
The simulations correspond to the parameters listed under the ``S-band" column in Tab.~\ref{tab:beamparam}.
\label{fig:S_perf}}
\end{figure}

We present, for the ``S-band" case of Tab.~\ref{tab:beamparam}, the evolution of the BFF over a frequency range $f\equiv\frac{\omega}{2\pi} \in [0.5, 3.5]$~THz as a function of 
drift distance from the DLW exit ($z_{drift}$) in Fig.~\ref{fig:S_perf}(a). The corresponding  longitudinal-density evolution appears in Fig.~\ref{fig:S_perf}(b).  For this set of parameters, 
10 microbunches are produced and a maximum bunching of $\widetilde{F}(\omega_1)\simeq 0.20 $ is obtained at the DLW fundamental mode's wavelength $\lambda_1 \simeq 382$~$\mu$m.
In addition, harmonics of the fundamental mode $f_{1,n}=nf_1$ are observed. For the selected DLW parameters and the corresponding thin dielectric layer,  only the 
fundamental mode significantly influences the bunch dynamics.

\begin{figure}[hhhh!!]
\centering
\includegraphics[width=0.46\textwidth]{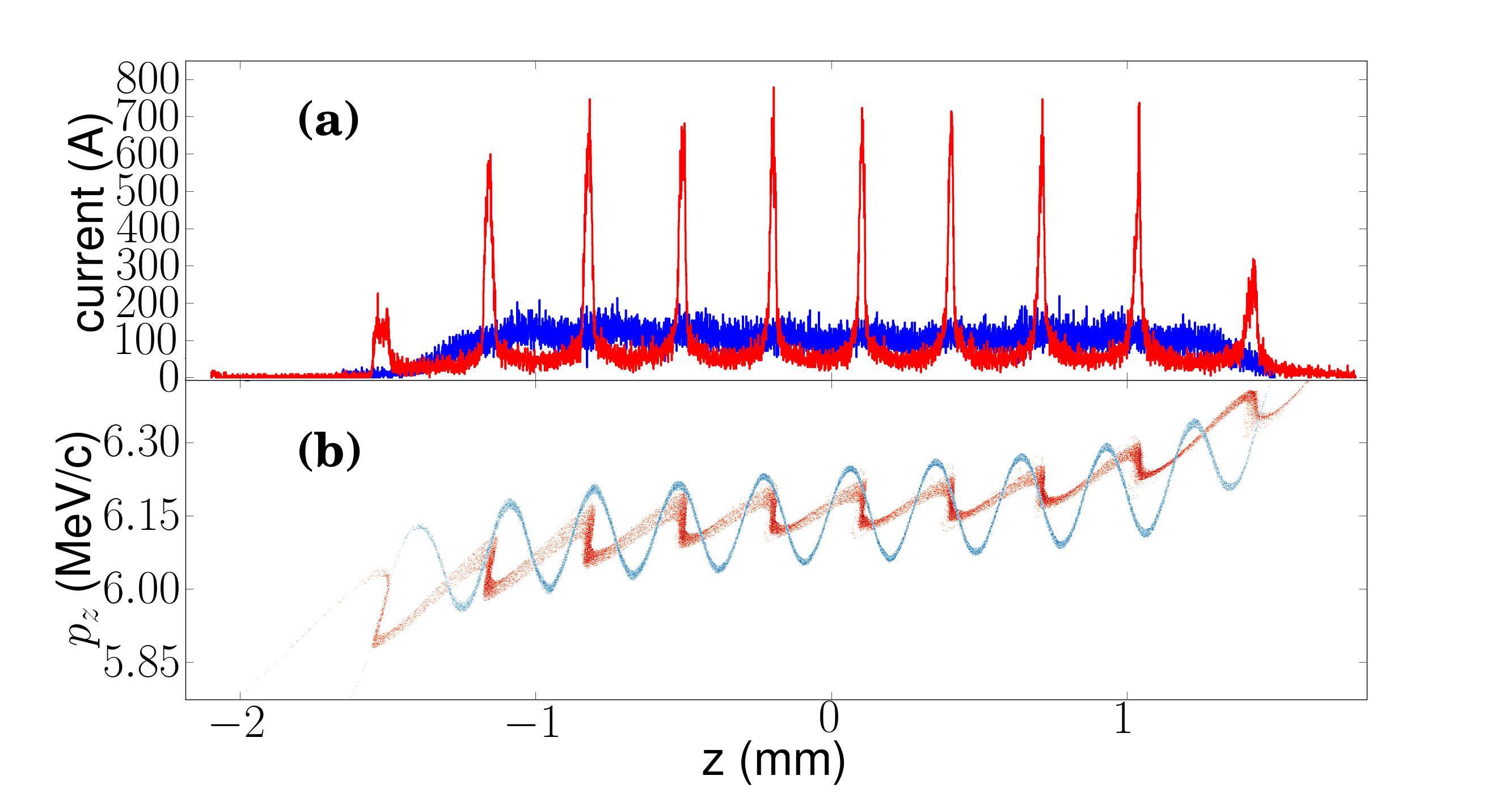} 
\includegraphics[width=0.46\textwidth]{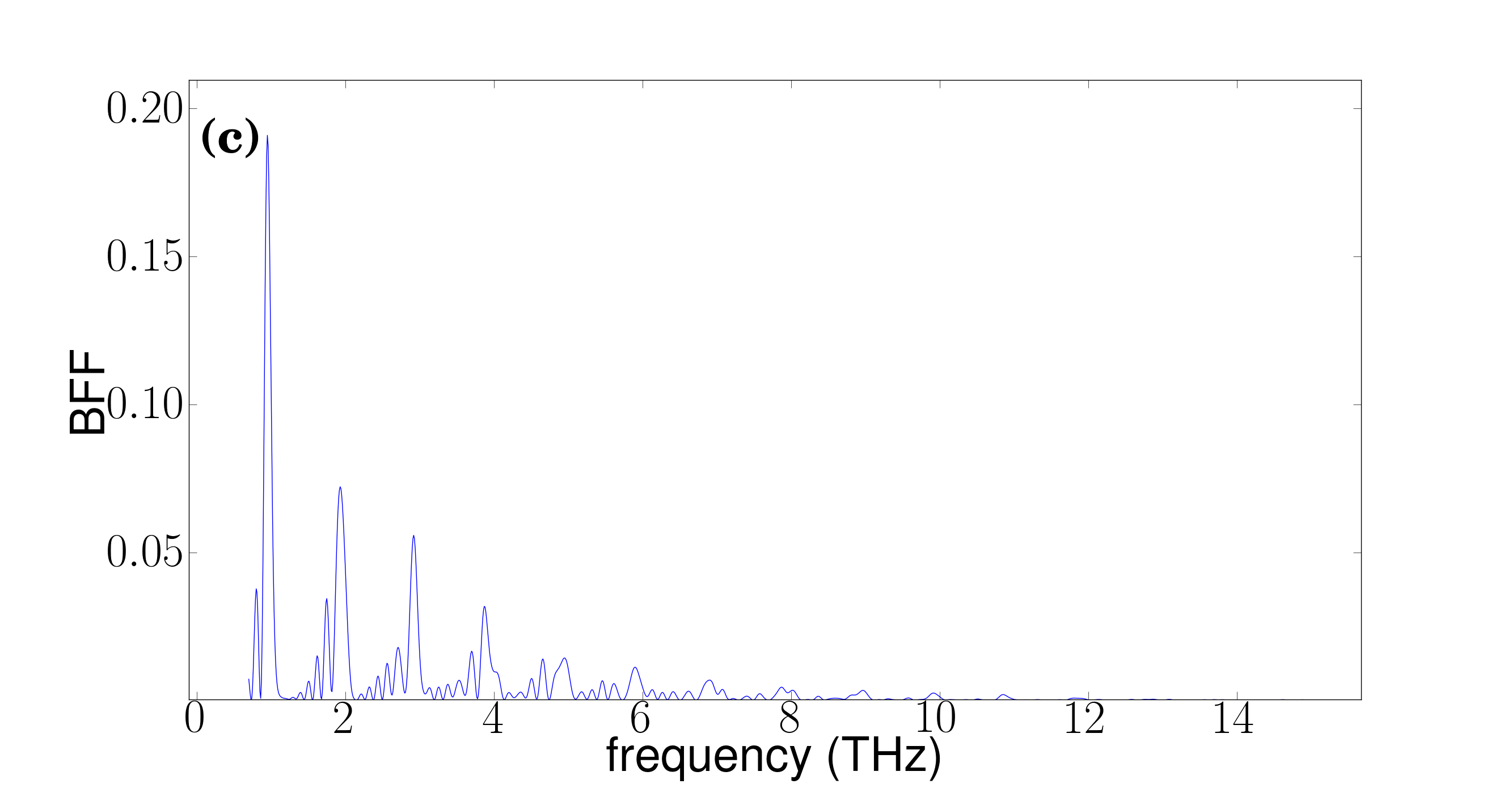}
\caption{Current profiles (a) and associated longitudinal phase spaces (LPS) (b) simulated at the entrance of the DLW structure (blue traces) and at the location of maximum bunching 
$z=1.3$~m from the photocathode. Bunch form factor (BFF) (c) obtained at $z=1.3$~m from the photocathode. 
The simulations correspond to the parameters listed under the ``S-band" column in Tab.~\ref{tab:beamparam}.
\label{fig:S_LPScurrentBFF}}
\end{figure}

The current and LPS distributions at the DLW entrance and at the location of maximum bunching (at $s\simeq 1.30$~m from the photocathode) appear in Fig.~\ref{fig:S_LPScurrentBFF}. 
Peak current on the order of 1~kA are achieved for a beam with  mean momentum of $\mean{p_z}\simeq 6.12$~MeV/c. The shortest current spike generated has an full-width 
half-max (fwhm) duration of $\sim 30$~fs. These 
results are comparable to the one experimentally obtained through wave-breaking in Ref.~\cite{renkai} albeit with a much higher contrast ratio~\cite{pietroLSC}. The origin of the 
non-uniform bunching  across the beam with peak-to-peak variation in the microbunch current is twofold. First, the slice-energy-spread positional variation along the bunch affects the shortest structure achievable at a given location. Second, the LPS prior to the DLW has initial correlations [as seen on the blue density plotted in Fig.~\ref{fig:S_LPScurrentBFF}(b)]  which affect the bunching uniformity across the microbunches. This latter initial correlation is also responsible for the apparent ``walk-off" feature (the microbunches spread apart from each others as they drift) of the microbunches visible in Fig.~\ref{fig:S_perf}(b). Figure~\ref{fig:S_LPScurrentBFF}(c) indicates strong harmonic content at the second and third harmonic frequencies of $f_1$ is also observed at the location of maximum bunching.

\begin{figure}[hhhh!!]
\centering
\includegraphics[width=0.46\textwidth]{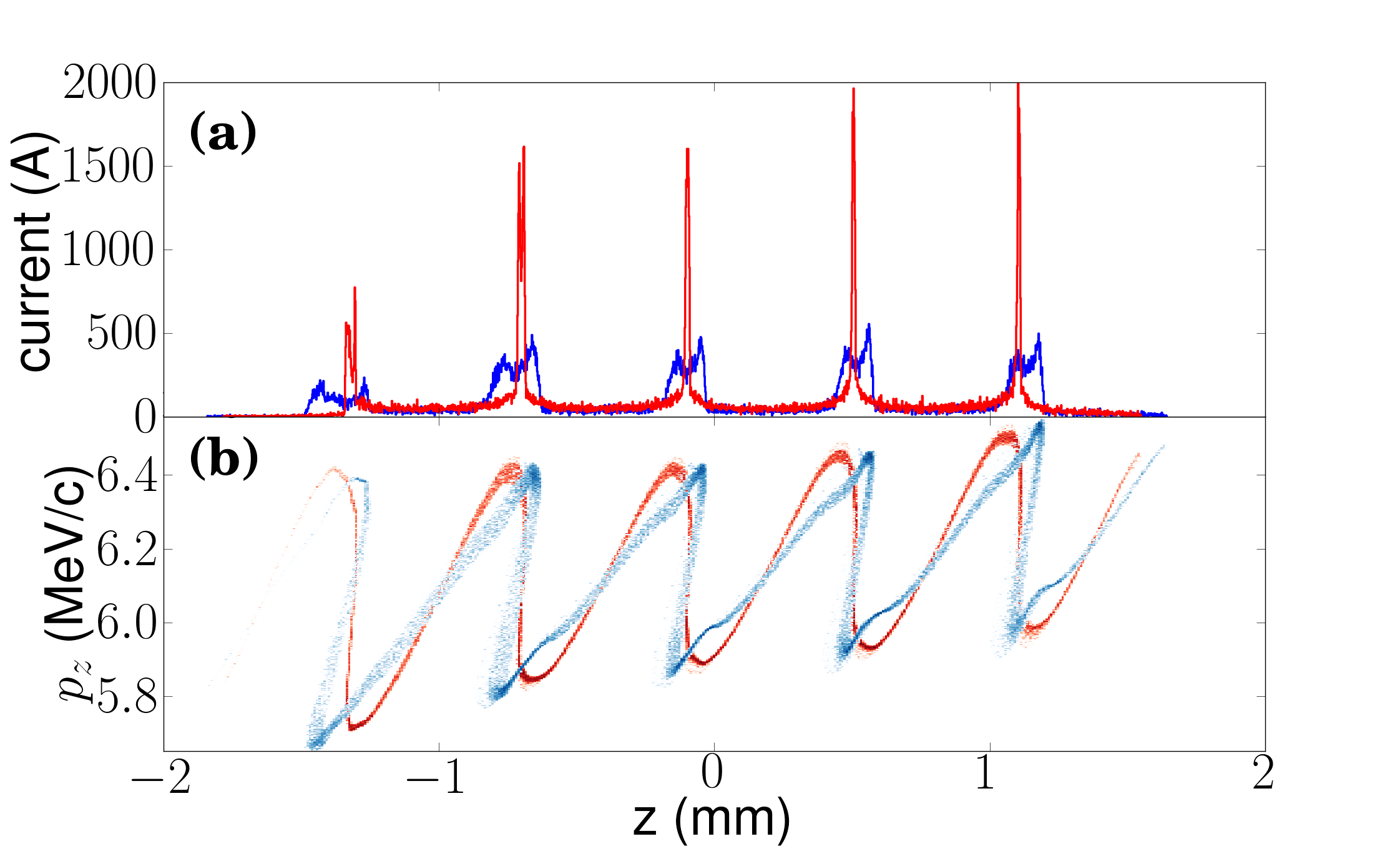}
\includegraphics[width=0.46\textwidth]{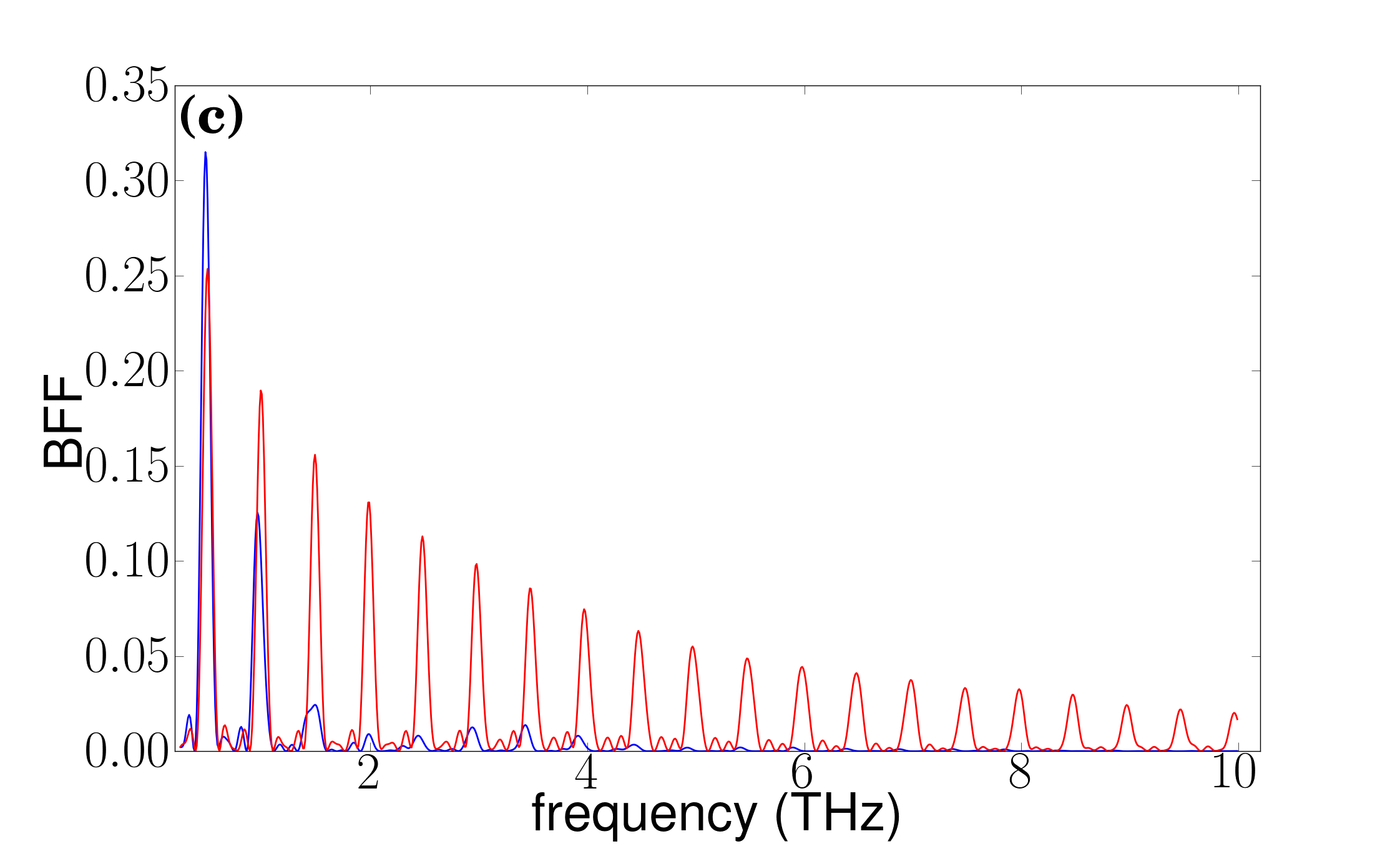}
\caption{Current profiles (a) and associated longitudinal phase spaces (LPS) (b) simulated at maximum compression 31~cm downstream of the DLW (red traces) and at the location of slight over-compression
52~cm downstream of the DLW. Bunch form factor (BFF) (c) obtained at the similar locations.
The simulations correspond to the parameters listed under the ``S-band" column in Tab.~\ref{tab:beamparam} with the exception of the geometric parameters of the DLW structure selected to be 
 $a= 350$~$\mu$m, and $b= 393$~$\mu$m.
\label{fig:S_LPScurrentBFF2}}
\end{figure}

Moreover, the higher harmonics are limited by the precision of the micro-bunch spacing within the bunch; a higher frequency DLW will lead to more micro bunches which
will be more limited by the initial correlated LPS.  We can investigate this feature by using a lower frequency structure of 500~GHz in the same
context of the 1~THz example illustrated above.  The current and LPS is shown in Fig.~\ref{fig:S_LPScurrentBFF2}(a,b) and associated BFF over the frequency range (0.25~THz, 10~THz),
is shown in Fig.~\ref{fig:S_LPScurrentBFF2}(c) for maximum compression (red trace).  The very strong higher harmonic content is notably due to the precise spacing of the 
microbunches.  Additionally, we may want to suppress higher harmonics or amplify the fundamental; this can easily be done by selecting a bunch which is under or over-compressed
such that the micro-bunches span a larger spatial extent; see Fig.~\ref{fig:S_LPScurrentBFF2}(a,b,c) blue trace. 

\begin{figure}[hhhh!!]
\centering
\includegraphics[width=0.49\textwidth]{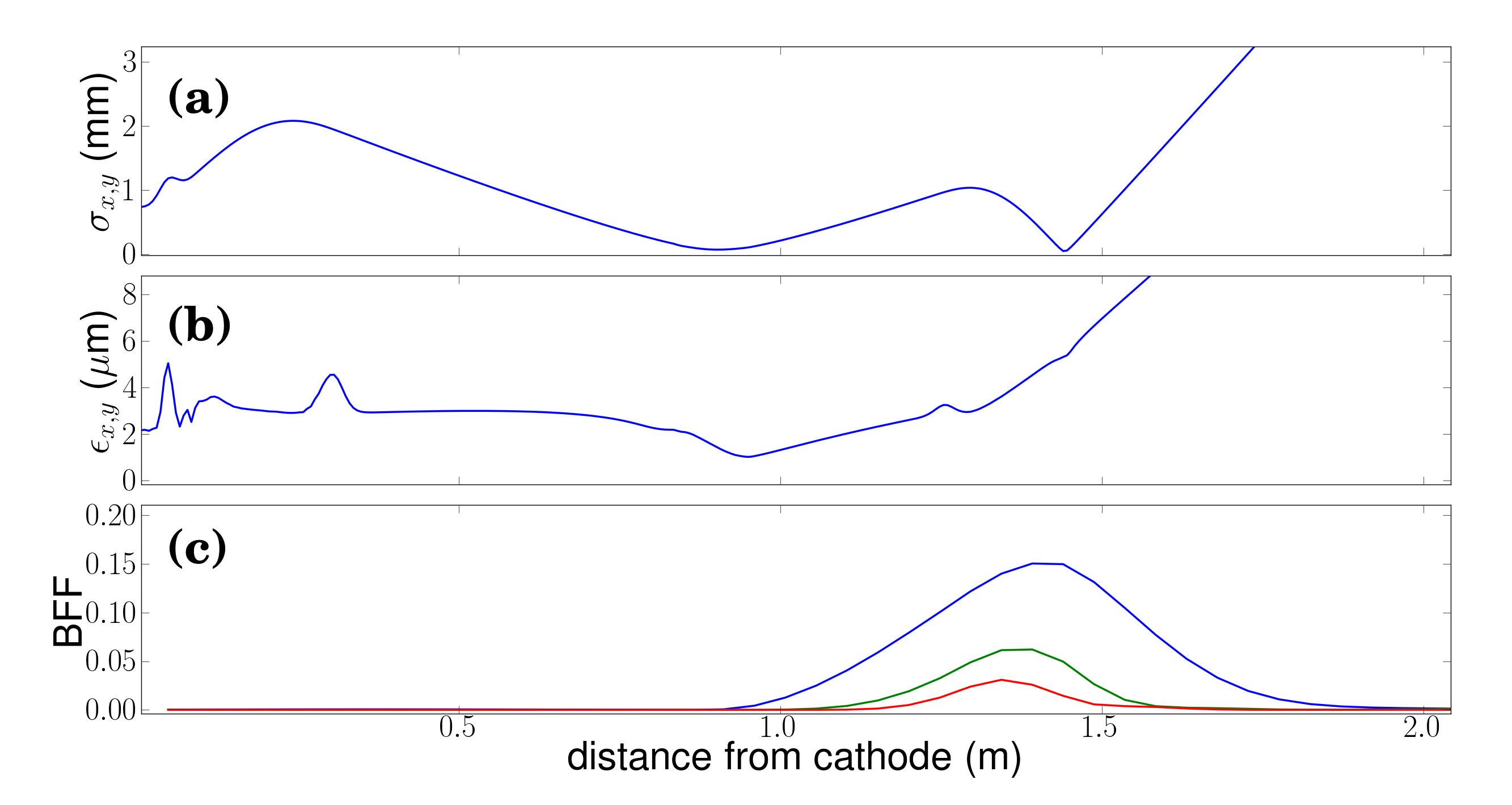} 
\caption{Transverse horizontal $\sigma_x$ and vertical $\sigma_y$ rms beam sizes (a), corresponding transverse emittances (b) and  bunch form factor (BFF) (b) evolution 
along the beamline. The BFF is evaluated at $f_1= 1$~THz (blue trace) and at the second  (green trace) and third (red trace) harmonics. The simulations correspond to the 
parameters listed under the ``S-band" column in Tab.~\ref{tab:beamparam}.
\label{fig:S_envelopeandBFF}}
\end{figure}

Finally, the evolution of the transverse beam sizes and emittance is respectively shown in Fig.~\ref{fig:S_envelopeandBFF}(a) and (b) for the case presented in Fig.~\ref{fig:S_LPScurrentBFF}. The 
addition of a second solenoid at $s\simeq 1.2$~m can transversely focus the beam down to $\sigma_x=\sigma_y\simeq 45$~$\mu$m at an axial location close to the maximum bunching; see  
Fig.~\ref{fig:S_envelopeandBFF}(c). The simulated small rms beam size confirms that the one-dimensional BFF approach adopted earlier can accurately be used to estimate the 
properties of radiation emitted at wavelengths $\lambda \gg \gamma^{-1}  \sigma_{x,y} \sim  5$~$\mu$m. It is therefore applicable to the THz regime. The small transverse size 
could also permit the use of a second DLW as a narrowband THz radiator as explored in Ref.~\cite{cook}. 

The location of maximum bunching depends primarily on the wakefield amplitude compared to the average bunch energy. Applying higher 
peak fields in the RF-gun leads to larger ballistic bunching lengths downstream of the DLW structure and vice versa. Alternatively, shorter bunching lengths can 
be achieved by decreasing the bunch length at the cost of more microbunches. To confirm the applicability of our concept to other configurations we carried 
a similar study as the one presented above for the case of an L-band RF gun.  

\begin{figure}[hhhh!!]
\centering
\includegraphics[width=0.46\textwidth]{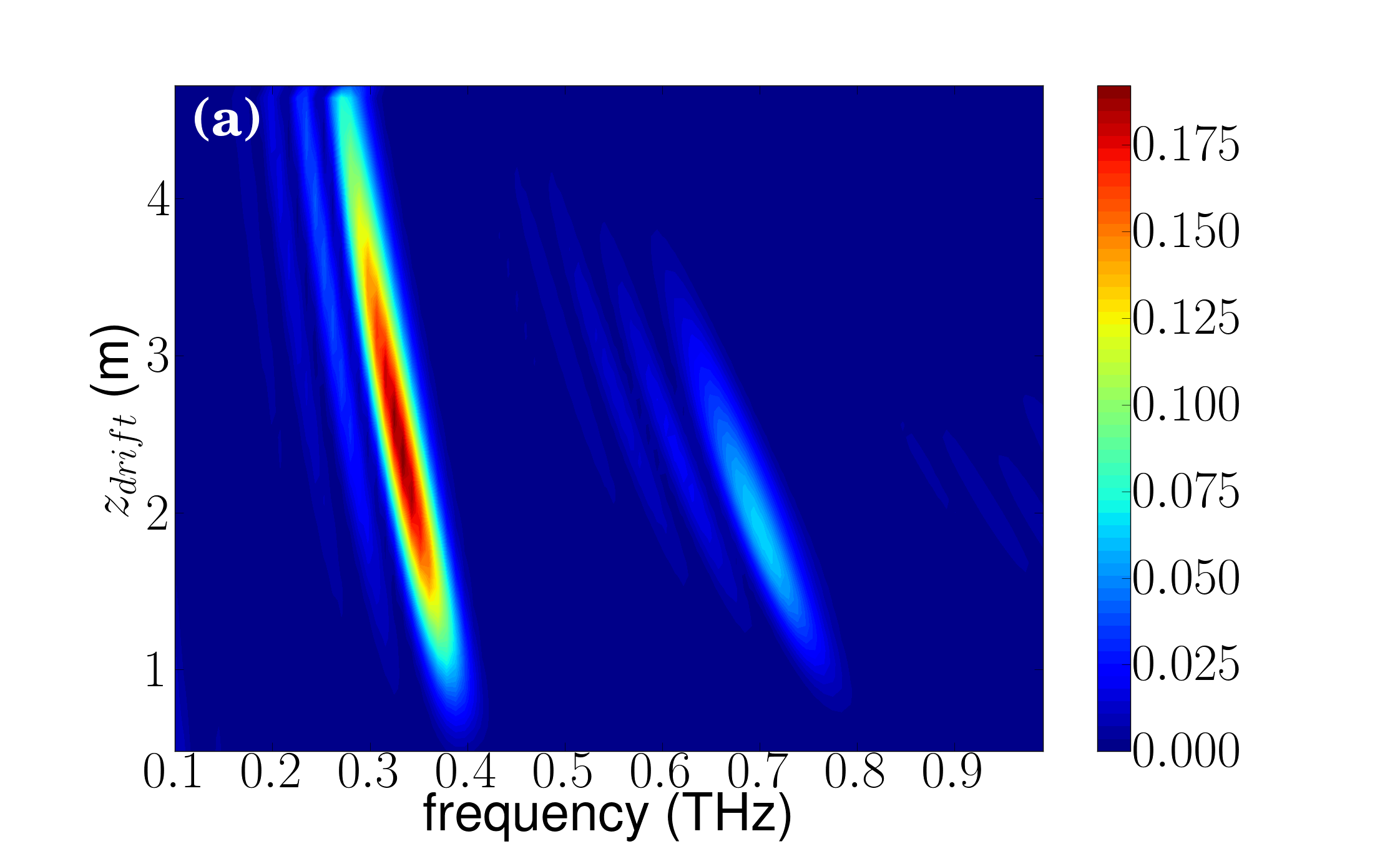}
\includegraphics[width=0.46\textwidth]{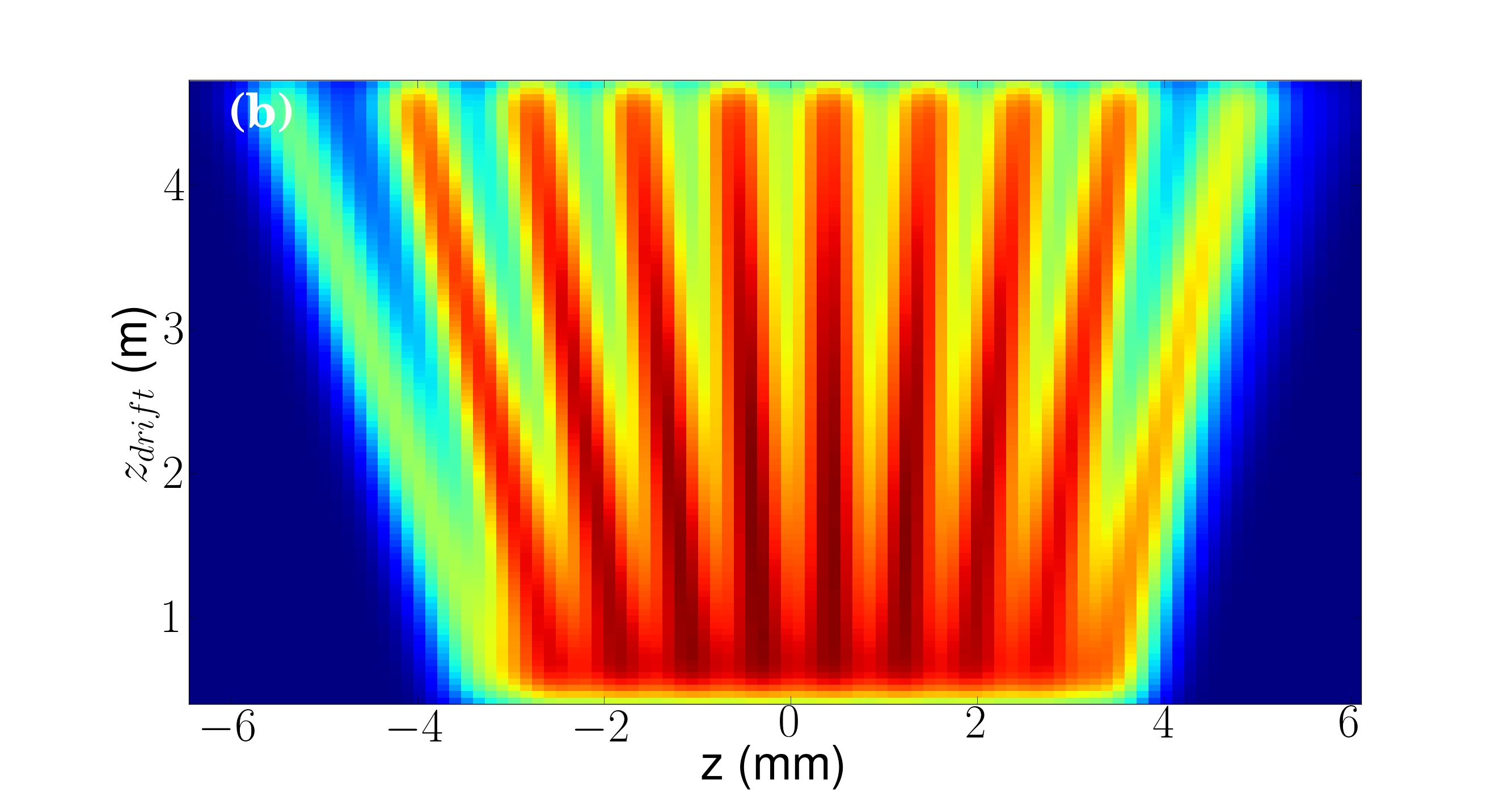}
\caption{Bunch form factor (BFF) (a) and bunch longitudinal density (b) evolution as a function of the drift length referenced with respect to the DLW exit.  
The simulations correspond to the parameters listed under the ``L-band" column in Tab.~\ref{tab:beamparam}. 
\label{fig:L_perf}}
\end{figure}

For this case we consider the setup available at the Fermilab's A0 photoinjector~\cite{carneiro} which incorporates a first-generation 
L-band gun used at the decommissioned Tesla-test facility at DESY~\cite{TTF1}. The gun is nested in three solenoidal lenses. An optimization similar 
to the one carried for the S-band case was conducted and the resulting operating parameters are displayed in Tab.~\ref{tab:beamparam} (``L-band" column).  
For completeness the BFF and longitudinal density evolution downstream of the DLW are shown in Fig.~\ref{fig:L_perf}. As in the S-band case we observe 
strong bunching at the DLW fundamental mode's frequency (in this case $\lambda_1\simeq 750$~$\mu$m as the DLW parameters are different). But in  contrast 
with  the S-band case the higher-harmonic content of the BFF are significantly suppressed. The change in the fundamental frequency as the bunch drift downstream of 
the DLW appear stronger than for the S-band case and is due to a more prominent  ``walk-off" effect due to the lower beam energy. 


\subsection{Passive Bunching and Shaping}

We now turn to another potential application of the scheme detailed in Section~\ref{sec:scheme} to bunch or shape an electron beam produced via photoemission 
from an RF gun (this corresponds  to the case when  $\sigma_z \lesssim \lambda_1$). \\

To illustrate our point, we consider the case of the L-band gun just discussed in the previous section and instead of using the DLW parameters 
of Tab.~\ref{tab:beamparam}, we consider a structure with inner radius $a=650$~$\mu$m to produce a global correlated energy spread as 
the fundamental-mode wavelength of the DLW becomes comparable to the bunch length. We use a 10-cm long DLW structure and note that due to the 
relatively large wavelengths required, the aperture becomes larger and the beam size requirement are significantly relaxed.  As mentioned 
earlier, the inherent nonlinear LPS distortion exhibits a correlation between the depleted energy location and tail that has the proper sign for 
compression via ballistic bunching.  

\begin{figure}[hhhh!!]
    \centering
    \includegraphics[width=0.47\textwidth]{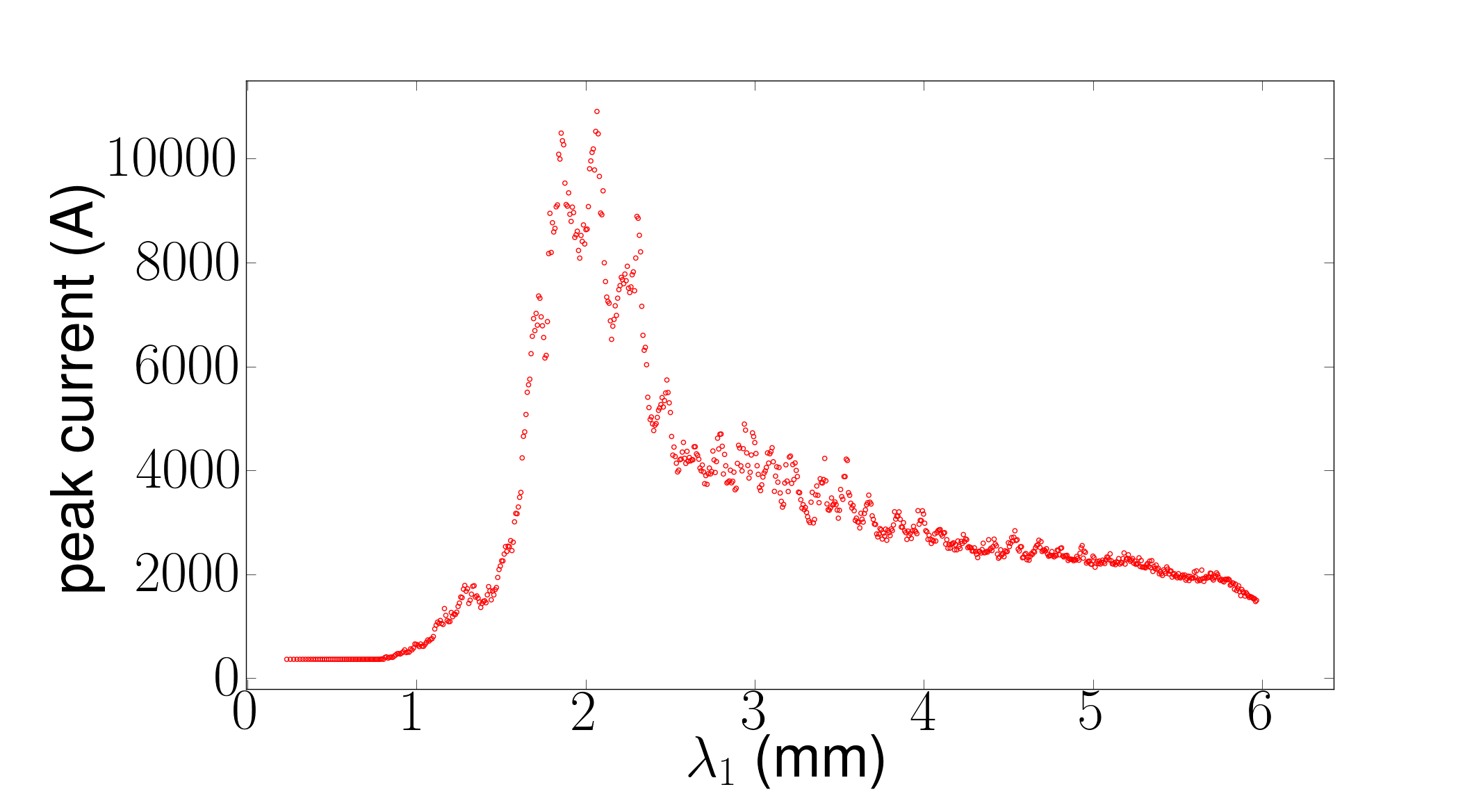}
    \caption{Maximum peak current as function of the fundamental-mode wavelength $\lambda_1$. The observed noise comes from numerical errors in precisely determining  
    the value of the the axial position where the peak current is maximized. These simulations are carried with the beam parameters summarized in Tab.~\ref{tab:beamparam} 
    ``L-band" column but for a DLW structure with inner radius $a=650$~$\mu$m. The fundamental-mode wavelength is varied by changing the structure outer radius $b$. 
    \label{fig:scanbc}}
\end{figure}

\begin{figure}[hhhh!!]
	\centering
	\includegraphics[width=0.48\textwidth]{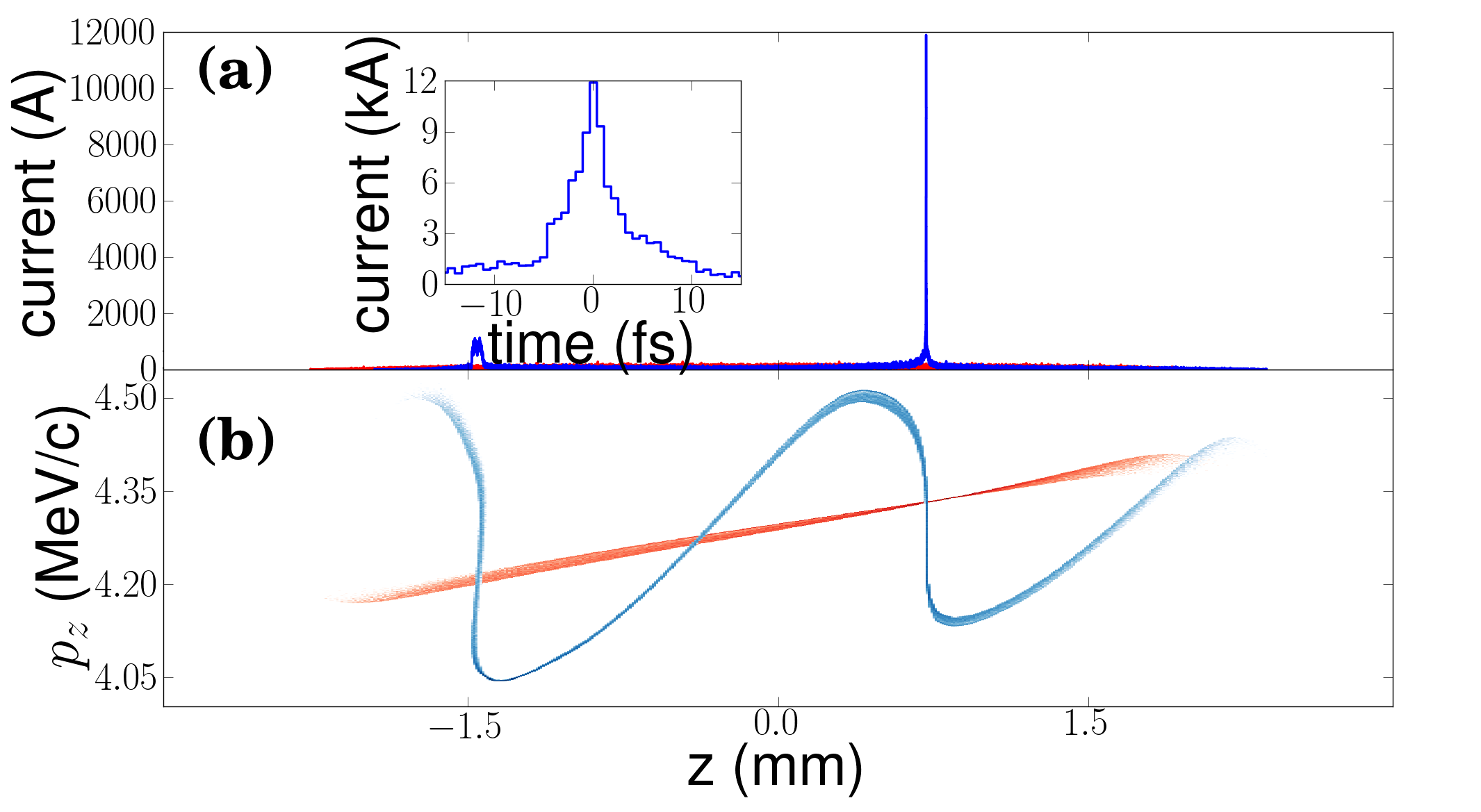}
	\caption{Current profiles (a) and longitudinal phase spaces (LPS) (b) at the entrance of the DLW structure (red traces) and at location of maximum compression (blue traces). 
	The simulation correspond to the case $\lambda=2.06$~mm in Fig.~\ref{fig:scanbc}. The inset in plot (a) corresponds to a zoom of plot (a) around the $\sim 12$-kA peak with 
	its origin of the temporal axis corresponding to $z=0.693$~mm in plot (a) axial coordinate. Maximum bunching, in this scenario occurs 43.9~cm downstream of the DLW.\label{fig:highpeakcur}}
\end{figure}

We exemplify this possibility by exploring the change in peak current downstream of a DLW structure with fixed inner radius $a=650$~$\mu$m as a function 
of the fundamental-mode wavelength. The mode's wavelength is varied with different dielectric thicknesses due to the relatively small impact on the bunching length
compared to changing $a$ directly. The 
results appear in Fig.~\ref{fig:scanbc} and indicate that a peak current of $\sim 12$~kA are attained when the fundamental-mode wavelength is $\sim  2.06$~mm
(corresponding to   $\lambda_1=0.49\sigma_z$). 
The latter wavelength 
corresponds to a structure with outer radius $b = 855$~$\mu$m (or dielectric thickness $\tau\equiv b-a=205$~$\mu$m). The associated current profiles 
and LPS appear in Fig.~\ref{fig:highpeakcur} and illustrates the role of the initial longitudinal emittance of the bunch before the DLW (i.e. the maximum 
peak current is achieved for an initial axial slice with the smallest slice energy spread.) In Fig.~\ref{fig:highpeakcur} only 7.1\% of the population resides 
within the current spike while the rest contributes to the formation of longitudinal tails. This low-current population of the bunch could in principle be reduced via dispersive scraping or by 
exploring some energy-transverse correlations in conjunction with transverse collimators.  Also, due to the relatively large inner radii needed to support wavelengths comparable to the bunch length, this technique can in principle easily be scaled to higher bunch charges. Finally, we note that the current profiles shown in Fig.~\ref{fig:highpeakcur} can actually find applications, 
e.g. to investigate wakefield effects in accelerating structures~\cite{asset} and in compact beam-driven acceleration schemes utilizing low-energy drive 
bunches. \\


\begin{figure}[hhhh!!]
    \centering
    \includegraphics[width=0.48\textwidth]{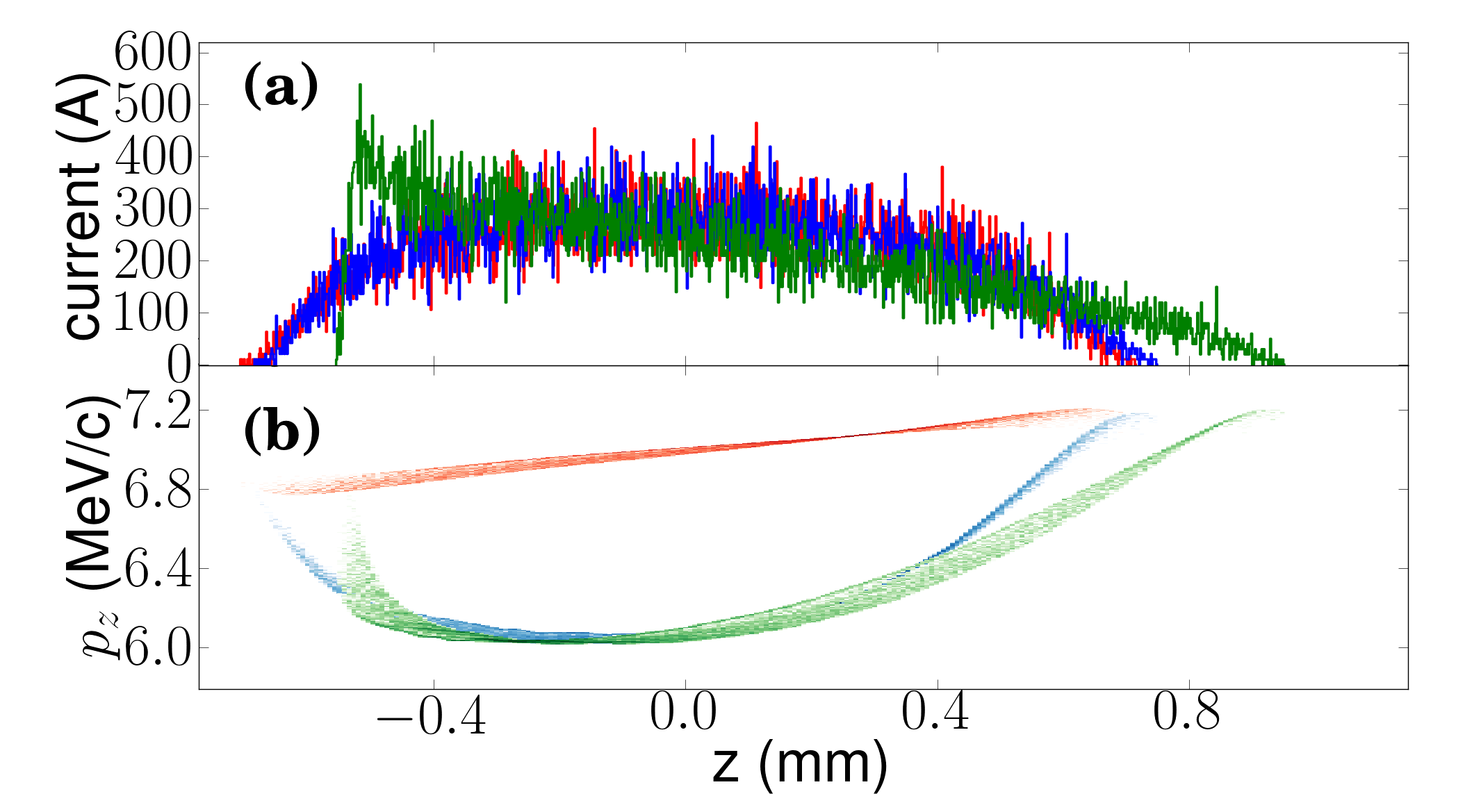}
    \caption{Current profiles (a) and longitudinal phase spaces (b) at the entrance (red traces)  and exit  (red traces) of the DLW structure and 0.2-m downstream of the structure ($s\simeq 0.54$~m from the photocathode surface) 
    where a quasi-linear current profile is achieved (green traces).
    \label{fig:rampevo}}
\end{figure}

As a final application we investigate the possibility to produce low-energy bunches with linearly-ramped current profiles.  This type of distribution is sought after to improve 
the transformer ratio -- the maximum accelerating wakefield over the decelerating field experienced by  the driving bunch --  in collinear beam-driven acceleration 
schemes~\cite{bane}.  We demonstrate that a standard distribution typically produced downstream of an RF gun can be transformed into a ramped bunch with 
quasi-linear dependency on $z$. We take the example of the S-band gun considered in Sec.~\ref{sec:train} and set $L/\lambda_1 \approx 1/2$ where
$L$ is the full longitudinal size of the bunch upstream of the DLW structure.  For these simulations, the axial-field amplitude at the cathode is set to $E_0=140$~MV/m. Such 
an increase (compared to the set of parameters displayed in Tab.~\ref{tab:beamparam}) was required to mitigate bunch lengthening. 
Figure~\ref{fig:rampevo} depicts the LPS evolution and associated current profiles associate to the bunch as it enters (red trace), exits (blue trace) the DLW and after a drift of 
0.2~m (green traces). The interplay of  the DLW field and longitudinal-space charge force results in the appearance of nonlinear correlations in the LPS. These nonlinearities 
provide some control over the current profile.

\begin{figure}[hhhh!!]
    \centering
    \includegraphics[width=0.48\textwidth]{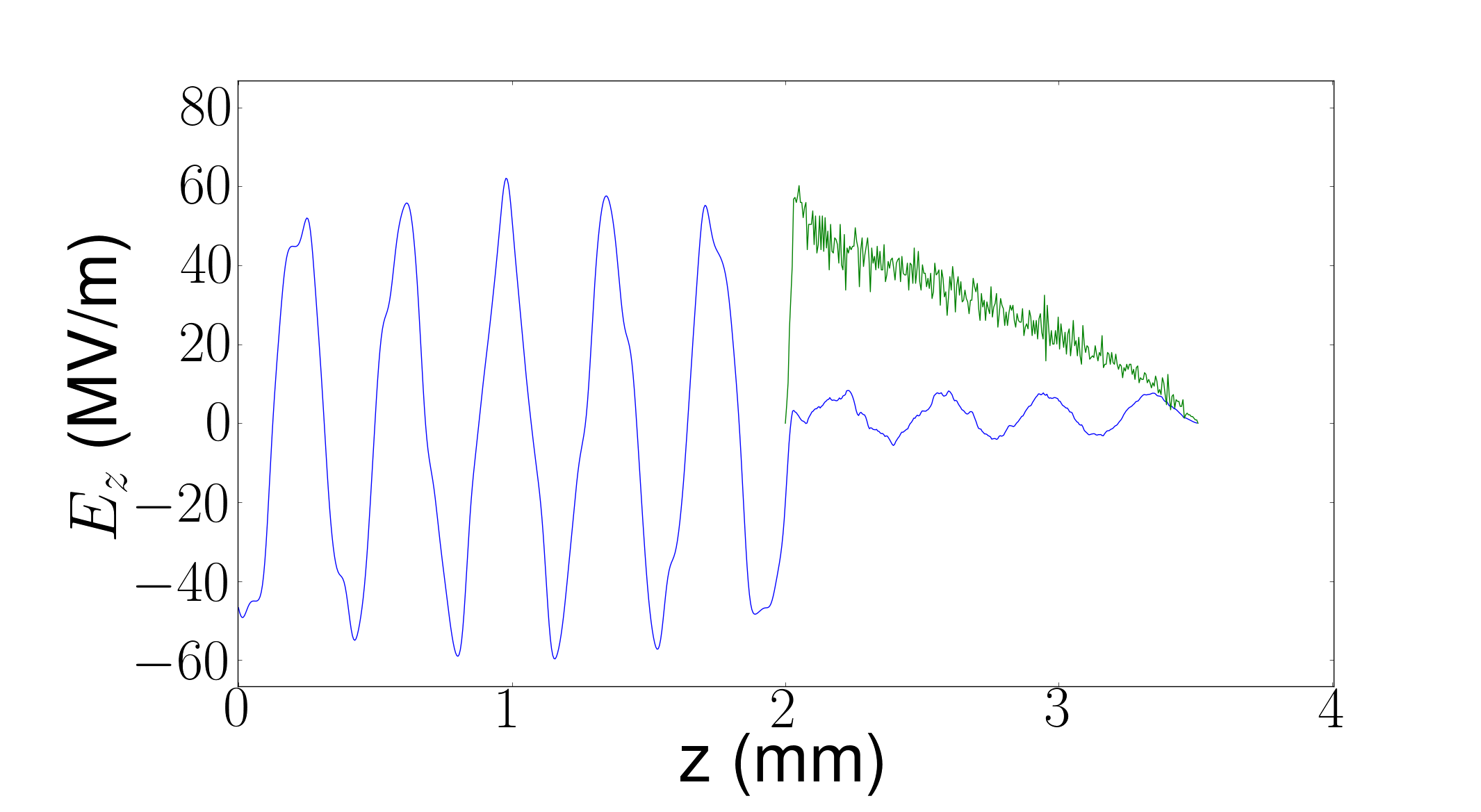}
    \caption{Longitudinal wakefield (blue trace) produced behind a bunch with the longitudinal distribution (green trace) identical to the one shown 
    in Fig.~\ref{fig:rampevo} [plot (a), green trace] for a bunch charge of 1~nC. The structure used 
    for the wakefield generation has the geometric parameters  $a=165$~$\mu$m,  $b=197$~$\mu$m and $ \varepsilon_r= 5.7$.
    \label{fig:RampWake}}
\end{figure}

To quantify the performance of the current profiles simulated in Fig.~\ref{fig:rampevo}(b, green trace), we compute the expected wakefield that such a bunch would 
produce in a  DLW structure optimized to sustain a large axial field. The structure's inner and outer radii are respectively $a=165$~$\mu$m, $b=197$~$\mu$m and the 
relative dielectric permittivity is kept to $ \varepsilon_r= 5.7$. The resulting wakefield behind the bunch has a peak accelerating field amplitude of $E^+\simeq 60$~MV/m; see 
Fig.~\ref{fig:RampWake}. The transformer ratio is numerically inferred as ${\cal R} \equiv |E_+/E_-|$ where  $E_-\simeq 8.2$~MV/m  is the maximum amplitude of the 
decelerating electric field within the electron bunch.  The achieved transformer ratio of ${\cal R}\simeq 7.3$ is comparable to the ideal ratio of ${\cal R}= n_p \pi \simeq 9.4$ 
predicted for an ideal linearly-ramped current profile (here $n_p\simeq 3$ is the number of mode wavelength comprised within the total bunch length)~\cite{bane}.  
Depending on the desired application, the photoinjector settings and DLW parameters could be adjusted to produce a ramped current profile after further acceleration 
in a subsequent linac. 

Finally, a finer control over the bunch shape could possibly be implemented using several DLW structures with properly selected fundamental-mode wavelengths. Such a 
multifrequency DLW approach would be an extension of the scheme described in Ref.~\cite{piotPRLshaping} to higher frequencies.

\section{SUMMARY}
In summary, we presented a relatively simple technique to bunch non-ultrarelativistic beams commonly produced by photoinjectors. The method was shown to 
support the generation of bunch train consisting of sub-picosecond microbunches. Alternatively, we demonstrated that a DLW with  a lower-frequency fundamental 
mode could act as a passive buncher and produce multi-kA bunches. In addition, we discuss the application of the technique to  form  bunches with linearly-ramped 
current profiles as needed to improve the transformer ratio in beam-driven advanced-acceleration techniques. One of the main advantages of the method is that it 
relies on the bunch interaction with its self-induced fields which are inherently synchronized: the technique is therefore not prone to temporal jitter. 

We expect the proposed method to find useful applications that span accelerator-based compact THz-radiation sources,  ultra-fast electron diffraction and in 
photoinjectors for short-wavelength linacs.

 It is also worth noting that the scheme could in principle be combined with other electron-emission process  (e.g. 
thermionic- or field-emission) but a detailed exploration is beyond the scope of the present study. 

Finally, other wakefield mechanisms, e.g., the use of a corrugated pipe~\cite{marcus2,corrugated} could provide an alternative to DLW and lead to similar results~\cite{PIERS}. Our 
selection of a DLW structure was mainly motivated by its manufacturing simplicity and wide use in advanced accelerator R\&D. \\

This work was supported by the Defense Threat Reduction Agency, Basic Research Award \# HDTRA1-10-1-0051, to Northern Illinois University and by the Department of Energy  contracts No. DE-FG02-08ER41532  and No. DE-SC0011831 with Northern Illinois University. PP is partially supported by DOE contract DE-AC02-07CH11359 to the Fermi research alliance LLC.


\begin{thebibliography}{99}   
%
\bibitem{zewai} A. H. Zewail and J. M. Thomas, {\em 4D Electron Microscopy: Imaging in Space and Time}, Imperial College Press, London, 2010.
\bibitem{ued} R. K. Li, P. Musumeci, H. A. Bender, N.S. Wilcox, M. Wu, J. Appl. Phys. {\bf 110} (7), 074512 (2011). 
\bibitem{gover} A. Gover, Phys. Rev. ST Accel. Beams {\bf 8}, 030701 (2005).
\bibitem{mueller} A.-S.  M\"uller, Rev. Accl. Sci. Tech., {\bf 03}, 165 (2010). 
\bibitem{aard1} A. Modena, Z. Najmudin, A.E. Dangor, C.E. Clayton, K.A. Marsh, C. Joshi, V. Malka, C.B. Darrow, C. Danson, D. Neely, and F.N. Walsh, Nature (London) {\bf 377}, 606 (1995).
\bibitem{aard} D.F. Gordon, A. Ting, T. Jones, B. HaÞzi, R.F. Hubbard, P. Sprangle, ``Particle-in-cell simulation of optical injector for plasma accelerators", in Proceedings of the 2003 Particle Accelerator Conference (PAC'03), 1846 (2003).
\bibitem{carlstenBC} B. E. Carlsten and S. M. Russel, Phys. Rev. E {\bf 53}, R2072-R2075 (1996). 
%
\bibitem{velo1} X. J. Wang, , X. Qiu, and I. Ben-Zvi,  Phys. Rev. E {\bf 54} R3121 (1996).
\bibitem{velo2} X. J. Wang, X. Y. Chang, Nucl. Instr. Meth. A {\bf 507} 310 (2003). 
\bibitem{velobunch} P. Piot, L. Carr, W. S. Graves, and H. Loos,  Phys. Rev. ST Accel. Beams {\bf 6}, 033503 (2003).
\bibitem{ferrario} M. Ferrario, et al. Phys. Rev. Lett. {\bf 104} 054801 (2010). 
%
\bibitem{luiten} T. van Oudheusden, P. L. E. M. Pasmans, S. B. van der Geer, M. J. de Loos, M. J. van der Wiel, and O. J. Luiten
Phys. Rev. Lett. {\bf 105}, 264801 (2010)
\bibitem{klaus} K. Fl\"ottmann, Nucl. Instr. Meth.  A {\bf  740}, 34 (2014).
\bibitem{Thzbuncher} C. Sung, S. Ya. Tochitsky, S. Reiche, J. B. Rosenzweig, C. Pellegrini, and C. Joshi, 
Phys. Rev. ST Accel. Beams {\bf 9}, 120703 (2006). 
\bibitem{franz} L. J.  Wong, A. Fallahi, and F. X. K\"artner, Optics Express, {\bf 21} (8), 9792 (2013). 
\bibitem{muggli} P. Muggli, V. Yakimenko, M. Babzien, E. Kallos, and K. P. Kusche, Phy. Rev. Lett. {\bf 101}, 054801 (2008).
\bibitem{yinebunch} Y.-E Sun, P. Piot, A. Johnson, A. H. Lumpkin, T. J. Maxwell, J. Ruan, R. Thurman-Keup, Phys. Rev. Lett. {\bf 105}, 234801 (2010).
\bibitem{piot} P. Piot, Y.-E Sun, T. J. Maxwell, J. Ruan, A. H. Lumpkin,  M. M. Rihaoui, and R. Thurman-Keup,  Appl. Phys. Lett. {\bf 98}, 261501 (2011)
\bibitem{ychuang} Y.-C. Huang, Int. Jour. Mod. Phys. B 21, 287 (2007).
\bibitem{boscolo} M. Bolosco, I. Boscolo, F. Castelli, S. Cialdi, M. Ferrario, V. Petrillo and C. Vaccarezza, Nucl. Instr. Meth.  A {\bf 577}, 409 (2007).
\bibitem{yuelinmod} Y. Li and K.-J. Kim , Appl. Phys. Lett. {\bf 92}, 014101 (2008).
%
\bibitem{renkai} P. Musumeci, R. K. Li, and A. Marinelli, Phys. Rev. Lett. {\bf 106}, 184801 (2011).
\bibitem{antipov4} S. Antipov, C. Jing, M. Fedurin, W. Gai, A. Kanareykin, K. Kusche, P. Schoessow, V. Yakimenko, and A. Zholents, Phys. Rev. Lett. {\bf 108}, 144801 (2012). 
%
\bibitem{buncher1} D. R. Hamilton, J. K. Knipp and J. B. Horner Kuper, MIT Radiation Laboratory Series, {\em Klystron and Microwave Triodes} (McGraw-Hill, New York, 1948).
\bibitem{buncher2} B. C. Yunn, ``Physics of th JLAB 350-keV photoinjector", in Proceedings of the 1999 Particle Accelerator Conference, PAC'99 (New York, 1999), 2453 (1999). 
%
\bibitem{rg} M.~Rosing, and W.~Gai, Phys. Rev. D {\bf {42}}, 1829 (1990).

%
\bibitem{chao} A. Chao, {\em Physics of Collective Instabilities in High-Energy Accelerators},  Wiley Series in Beams \& Accelerator Technologies, John Wiley and Sons (1993). 
\bibitem{stupakov} G. V. Stupakov, {\em Wake and Impedance},  report SLAC-PUB-8683, unpublished report available from the Stanford Linear Accelerator Center  (2000). 
\bibitem{wilson} P. B. Wilson, {\em Introduction to wake fields and wake potentials},  reports SLAC-PUB-4547 and SLAC/AP-66, unpublished reports available from the 
Stanford Linear Accelerator Center  (1989). 
%
\bibitem{antipov1} S. Antipov, C. Jing, P. Schoessow, A. Kanareykin, B. Jiang, V. Yakimenko, A. Zholents, and W. Gai, AIP Conf. Proc. {\bf 1507}, 421 (2012). 
\bibitem{antipov2} S. Antipov, M. Babzien, C. Jing, M. Fedurin, W. Gai, A. Kanareykin, K. Kusche, V. Yakimenko, and A. Zholents, Phys. Rev. Lett. {\bf 111}, 134802 (2013)
\bibitem{antipov3} S. Antipov, C. Jing, P. Schoessow, A. Kanareykin, B. Jiang, V. Yakimenko, A. Zholents, and W. Gai, Rev. Sci. Instrum. {\bf 84}, 022706 (2013)
\bibitem{huang} Z. Huang, D. Dowell, P. Emma, C. Limborg-Duprey, ``Uncorrelated Energy Spread and Longitudinal Emittance of A Photoinjector Beam", in Proceedings of the Particle Accelerator Conference, 2005. PAC 2005, 3570 (2005)
\bibitem{huening} M. H\"uning and H. Schlarb, ÒMeasurements of the beam energy spread in the TTF photoinjectorÓ, in Proceedings of the 2003 Particle Accelerator Conference (PAC03), 2074 (2003).
\bibitem{astra}  K. Fl\"ottmann, {\em {\sc astra}: A space charge algorithm, User's Manual}, available from the world wide web at  http://www.desy.de/$\sim$mpyflo/AstraDokumentation (unpublished).
\bibitem{dohlus} M. Dohlus, K. Fl\"ottmann, and C. Henning, ``Fast Particle Tracking With Wake Fields", preprint arXiv:1201.5270 [physics.acc-ph] (2012); also report DESY 12-02 (available from DESY Hamburg, Germany). 
\bibitem{saxon} J. S. Nodvick and D. S. Saxon, {Phys. Rev.} {\bf 96}, 180 (1954).
\bibitem{saldin}  E. Saldin, E. Schneidmiller, and M. Yurkov, Nucl. Instrum. Methods Phys. Res., Sect. A {\bf 539}, 499 (2005).
\bibitem{slacgun} D. T. Palmer, R. H. Miller, H. Winick, X.J. Wang,  K. Batchelor, M. Woodle, and I. Ben-Zvi, ``Microwave measurements of the BNL/SLAC/UCLA 1.6-cell
photocathode RF gun", in Proceedings of the 1995 Particle Accelerator Conference, PAC'95 (Dallas, TX, 1995), 982 (1996). 
\bibitem{pitz} B. Dwersteg, et al.,   Nucl. Instr. Meth.  A {\bf  393}, 93 (1997).
\bibitem{genOpt} M. Borland and H. Shang, {\tt geneticOptizer}, private communication (2009).
\bibitem{pietroLSC} P. Musumeci, R. K. Li, K. G. Roberts, and E. Chiadroni, Phys. Rev. ST Accel. Beams {\bf 16}, 100701 (2013). 
\bibitem{cook} A. M. Cook, R. Tikhoplav, S. Y. Tochitsky, G. Travish, O. B. Williams, and J. B. Rosenzweig, Phys. Rev. Lett. {\bf 103}, 095003 (2009).
\bibitem{carneiro} J.-P. Carneiro, N. Barov, H. Edwards, M. Fitch, W. Hartung, K. Floettmann, S. Schreiber, and M. Ferrario, Phys. Rev. ST Accel. Beams {\bf  8}, 040101 (2005).
\bibitem{TTF1} J. Andruszkow, et al. [TESLA Collaboration], Phys Rev Lett {\bf 85}, 3825 (2000). 
\bibitem{asset} C. Adolphsen et al. Next Linear Collider (NLC) Newsletter, {\bf 1}, No. 2, August 2000 unpublished report available from the Stanford Linear Accelerator Center (2000). 
\bibitem{bane} K. L. F. Bane, P. Chen, P. B. Wilson, ``On collinear wakefield acceleration", SLAC-PUB-3662 (1985).
\bibitem{piotPRLshaping} P. Piot, C. Behrens, C. Gerth, M. Dohlus, F. Lemery, D. Mihalcea, P. Stoltz, M. Vogt, Phys. Rev. Lett. {\bf 108}, 034801 (2012).
\bibitem{marcus2} M. H\"uning, H. Schlarb, P. Schm\"user, and M. Timm,  Phys. Rev. Lett. {\bf 88}, 074802 (2002). 
\bibitem{corrugated} P. Emma, M. Venturini, K. L. F. Bane, G. Stupakov, H.-S. Kang, M. S. Chae, J. Hong, C.-K. Min, H. Yang, T. Ha, W. W. Lee, C. D. Park, S. J. Park, and I. S. Ko, 
Phys. Rev. Lett. {\bf 112}, 034801 (2014). 
\bibitem{PIERS} G.  Mishra and G. Sharma, Progress In Electromagnetics Research M {\bf 36}, 47 (2014). 
\end{thebibliography}
\end{document}